\documentclass[pra,reprint,superscriptaddress,floatfix]{revtex4-1}
\usepackage{appendix}
\usepackage{graphicx} 
\usepackage{amsmath}
\DeclareMathOperator{\Tr}{Tr}
\usepackage{amssymb}
\usepackage{braket}

\usepackage{hyperref}
\hypersetup{
    colorlinks=true,
    linkcolor=red,
    citecolor=black,
    urlcolor=black
}

\begin{document}

\newcommand{\erf}{\text{erf}}

\title{Local Harmonic Approximation to Quantum Mean Force Gibbs State}

\author{Prem Kumar}
\email{premkr@imsc.res.in} 
\email{prem3141592@gmail.com}
\affiliation{Optics and Quantum Information Group, The Institute of Mathematical Sciences, C.I.T. Campus, Taramani, Chennai 600113, India.}
\affiliation{Homi Bhabha National Institute, Training School Complex, Anushakti Nagar, Mumbai 400094, India.}

\begin{abstract}
    When the strength of interaction between a quantum system and bath is non-negligible, the equilibrium state can deviate from the Gibbs state. But the expression of such a mean force Gibbs state in an arbitrary parameter regime is unknown and is numerically challenging to determine. In this work, we first review the local harmonic approximation to this problem [Maier et al., Phys. Rev. E 81, 021107 (2010)], which can accurately determine the mean force Gibbs state when either the system-bath coupling or the temperature is large, or when the third and higher derivatives of the potential are small compared to certain system-bath specific parameters. In the appropriate limit, we show that the local harmonic approximation reduces to the ultra-strong coupling and high temperature results recently derived in the literature. After deriving an estimate for the error induced by this method, we apply it to study some systems, like a quartic oscillator and a particle in a quartic double-well potential. We also apply this method to analyze the proton tunneling problem in a DNA recently studied in literature [Slocombe et al., Comm. Phys., vol. 5, no. 1, p. 109, 2022], where our results suggest the equilibrium value of the probability of mutation to be orders of magnitude lower than the steady state value obtained there ($10^{-8}$ vs $10^{-4}$).
\end{abstract}

\maketitle

\section{Introduction}

Continuous variable (CV) quantum systems have been immensely studied in the context of fields like quantum information processing, communication and computation \cite{braunstein2005quantum, Ralph2006Quantum, Djordjevic2022Hybrid, Inoue2022Toward, Yang2015Electromagnetically, Gross2011Atomic, Huang2019Generation}, quantum resource theories and thermodynamics \cite{Takagi2018Convex, Ferrari2020Asymptotic, Narasimhachar2019Thermodynamic}, quantum technology \cite{Hoskinson2008Quantum, Ferreira1997Tunnelling}, quantum chaos \cite{Monteoliva2000Decoherence,  B.C.Bag1998A, Jt1992Semiclassical, Chakrabarty2019Out} and open quantum system \cite{Liu2007Non-markovian, Vasile2011Quantifying, Hoerhammer2007Environment-induced, ruostekoski1998bose, fisher1985dissipative, grabert1985quantum, topaler1994quantum, Oliveira2006Quantum-classical}. Thermodynamic equilibrium properties of these CV systems are hence of great interest in many of these studies and applications. But when the strength of interaction between a quantum system and bath is non-negligible, the equilibrium state can deviate from the Gibbs state and is given by the mean force Gibbs state (MFGS)\cite{cresser2021weak}, defined as,
\begin{align}\label{eqn:definition of MFGS}
	\rho = Tr_{R}\left[ \frac{e^{-\beta H_{tot}}}{Z}\right]
\end{align}
where $H_{tot}$ is the full system-bath Hamiltonian and the partial trace is taken over the bath degrees of freedom. The MFGS has found several applications in fields like strong coupling quantum thermodynamics \cite{seifert2016first, philbin2016thermal, jarzynski2004nonequilibrium, jarzynski2004nonequilibrium, campisi2009fluctuation} and quantum chemistry \cite{allen2006molecular, maksimiak2003molecular, roux1999implicit, roux1995calculation} and is relevant to many other problems in quantum chemistry and biology.

For example, consider a particle in a double well (DW) potential \cite{ruostekoski1998bose, gillan1987quantum, liu2021understanding, fisher1985dissipative, grabert1985quantum, topaler1994quantum} interacting with a bosonic bath which finds application, for example, in quantum chemistry and biology to study processes like chemical reactions occurring in the presence of a solvent \cite{slocombe2022open, slocombe2022quantum, zhang2020proton, bothma2010role}. Finding an analytical expression MFGS for such a system is hence of interest. But most of these physical processes happen in the intermediate coupling regime, where an analytical expression for MFGS is not available. By intermediate coupling regime, we mean the regime in which the bath-induced timescales on the system are of the same order as the timescales associated with the bath itself (e.g., $\hbar/k_B T$) \cite{BRE02}.

Analytical expressions for MFGS in weak and ultra-strong coupling (USC) limits were derived recently \cite{cresser2021weak}. Further, corrections to the USC regime MFGS have been evaluated using dynamics \cite{trushechkin2022quantum} and perturbative expansion methods \cite{latune2021steady}. Additionally, high temperature expansion for the MFGS was also derived recently \cite{timofeev2022hamiltonian, gelzinis2020analytical}.

If the CV system is a harmonic oscillator (HO), then the expression for the MFGS is known analytically \cite{grabert1988quantum, weiss2012quantum}. But for a particle in an arbitrary 1D potential $V(q)$, an analytical expression for MFGS in the intermediate coupling regime remains a challenge.
Recently, two numerically exact approaches, “Time Evolving matrix Product Operator” (TEMPO) \cite{chiu2022numerical, strathearn2018efficient, strathearn2020modelling, feynman2000theory, makri1995tensor1, makri1995tensor2} and ``The hierarchical equations of motion" (HEOM) \cite{tanimura2020numerically, liu2021understanding, zhang2020proton}, have made these MFGS and steady state calculations much more computationally efficient. But for a particle in an arbitrary 1D potential $V(q)$, the system is infinite dimensional, and a typical numerical technique proceeds by truncating the system in the energy eigenbasis \cite{topaler1993system}. In the intermediate coupling regime, this is an issue because the effective dimension of the system increases with the coupling, causing the calculation to become computationally very expensive.

Here, we first review the `local harmonic approximation' (LHA) to MFGS, for a particle in a 1D potential $V(q)$ interacting with a non-interacting bosonic bath \cite{PhysRevE.81.021107}. This method is accurate when either the system-bath coupling or the temperature is large, or when the third and higher derivatives of the potential are small compared to certain system-bath specific parameters.

We find that, in appropriate limits, LHA reduces to the ultra-strong coupling and high temperature results recently derived in literature \cite{cresser2021weak, timofeev2022hamiltonian}.
After providing a re-derivation of LHA, we estimate the error associated with it. This error estimate helps us apply LHA to several problems that may be of physical interest, which include the quartic oscillator and a particle in a quartic double-well potential. We also apply LHA to the problem of determining the equilibrium value of the probability of mutation in a DNA, which we find to be orders of magnitude lower than what was estimated in earlier research \cite{slocombe2022open}. Finally, we investigate and confirm the validity of the LHA in the non-intuitive regime of inverted local potentials.


In \autoref{sec_lha_history}, we provide a brief history of LHA.
In \autoref{sec:motivation}, after re-deriving LHA, we estimate the error associated with it. In \autoref{sec:application}, we use LHA to study some CV open quantum systems, like a quartic oscillator and a quartic double-well potential.
\autoref{sec: proton tunneling in DNA} provides an application of LHA to the proton tunneling problem in DNA that has been recently investigated by Slocombe et al. \cite{slocombe2022open, slocombe2022quantum}. \autoref{sec: Conclusion} provides conclusion and discussions.

\section{A brief history of LHA}
\label{sec_lha_history}

The basic motivation for LHA rests on the fact that since the expression for MFGS is exactly known when the system is a harmonic oscillator, we can use this to perform a suitable harmonic approximation at every point on the potential of a system which is actually \textit{anharmonic}. Since, in the path integral approach, paths arbitrarily far away from the starting and ending points contribute less to the final result, the path integral only `sees' local features of the potential, which, in many situations, might actually be well approximated by some harmonic potential.

Intuitively, there are two length scales at play here. One dictates the distance at which the potential deviates from the harmonic form. Let us call this lengthscale $L_1$. The other dictates the standard deviation of all paths (with given starting and ending points) that have non-negligible contribution to the path integral. Let us call this lengthscale $L_2$. If $L_1 \gg L_2$, LHA will be applicable. It turns out that increasing temperature and system-environment coupling decrease $L_2$, hence increasing the applicability of LHA.

The core idea behind LHA has been noted before by several authors \cite{feynman1965path, 10.1063/1.1676560, JORISH1975378, HELLSING1985303, PhysRevE.81.021107, Ankerhold_2003, Deffner_2011, PhysRevE.84.031110, Coffey_2007, PhysRevE.78.031114, B614554J}.
In the classic textbook, Feynman et al. derived the partition function of a particle (isolated from any environment) in an arbitrary potential in the large temperature limit \cite{feynman1965path}. Such zero coupling limit, high temperature Gibbs state results were later reported by several other authors \cite{10.1063/1.1676560, JORISH1975378, HELLSING1985303}.

The open quantum system version of the problem was then studied by several authors in different limits, like high temperature, system-environment coupling, and the smooth function limit \cite{PhysRevE.81.021107, Ankerhold_2003, Deffner_2011, PhysRevE.84.031110, Coffey_2007, PhysRevE.78.031114, B614554J}. In the present work, we build upon this history by first providing a consolidated re-derivation of the LHA formalism, and then using it as a foundation to study some non-intuitive features of this approximation, like its validity even when the second derivative of the potential is negative, and as a result, the local harmonic potential that approximates the \textit{anharmonic} potential at that point is \textit{inverted}.

\section{LHA and an Estimate of its Error}\label{sec:motivation}

We consider a particle of mass $m$ in an arbitrary potential $V(q)$, linearly coupled to a bath of harmonic oscillators. This system is described by the  Hamiltonian,
\begin{equation}\label{CL_hamiltonia}
	H=\frac{p^2}{2 m}+V(q) \\ +\sum_k\left[\frac{p_k^2}{2 m_k}+\frac{1}{2} m_k \omega_k^2\left(q_k-\frac{c_k q}{m_k \omega_k^2}\right)^2\right],
\end{equation}
where $\omega_k$, $m_k$, and $c_k$ are the frequency, mass, and coupling strength of the $k$th oscillator, respectively. For simplicity, we assume a one-dimensional system, though our results are easily generalized to higher dimensions. Here and throughout, we assume $\hbar = 1$.
Throughout this article, we write operator matrix elements in position basis as  $O(q,\eta) = \bra{q+\eta}\Hat{O}\ket{q-\eta}$. Diagonal elements are obtained when $\eta=0$ and will be denoted by $O(q) =O(q,0)$.

Upon integrating out the bath degrees of freedom, the MFGS for the particle can be expressed as a Euclidean path integral \cite{weiss2012quantum},
\begin{equation}\label{eqn:general path integral}
	\rho(q,\eta) = Z_R^{-1} \int_{q+\eta}^{q-\eta} \mathcal{D}q(\tau) e^{-S_0[q(\tau)] - \Phi[q(\tau)]}
\end{equation}
where, $\int_x^y \mathcal{D}q(\tau)$ signifies $q(0) = x$ and $q(\beta) = y$ in the functional integral. The reduced partition function is the ratio of bath and system+bath partition functions, $Z_R = Z/Z_B$, and
\begin{align}\label{eqn:free hamiltonian action}
	S_0[q(\tau)] = &\int_0^\beta d\tau \left( \frac{1}{2} m \dot{q}(\tau)^2 + V\left(q(\tau)\right) \right)  \\ \label{eqn:IP}
	\Phi[q(t)]=& \frac{1}{4}\int_0^\beta d\tau \int_0^\beta d\tau' K\left( \tau - \tau' \right) \left(q(\tau) - q(\tau')\right)^2  
\end{align}
are the action for the free particle and the influence action, respectively. The bath correlation function is defined as,
\begin{align}
	K(\tau) &= \int_0^\infty d\omega J(\omega) \frac{\cosh{[\omega(\beta/2 - \tau)]}}{\sinh{[\omega \beta/2]}}\label{eqn:influence function cosh form}
\end{align}
in terms of the spectral density
\begin{equation}\label{eqn:spectral density general form}
	J(\omega)= \alpha\sum _k \frac{c_k ^2}{2 m_k \omega_k } \delta(\omega - \omega_k),
\end{equation}
where we have introduced the dimensionless number $\alpha$ to characterize the strength of the coupling.

At large system-bath coupling, the influence action [\autoref{eqn:IP}] dominates over the free action, such that amplitudes for paths with $q(\tau)\ne q(\tau')$ are exponentially small in the path integral \autoref{eqn:general path integral}. In the limit $\alpha \to \infty$, the ultra-strong coupling (USC) limit, the amplitudes for these paths vanish and only the constant path contributes, $q(\tau)= q(\tau')=q$ with $\dot{q}(\tau)=0$. Similarly, at high temperature (but arbitrary coupling), the path length $\beta = 1/k_B T$ becomes small, so the kinetic term in the free action, $\propto \dot{q}(\tau)^2$, increasingly suppresses amplitudes for paths that deviate significantly from $q(\tau) = q$. In the limit $T \to \infty$ (zero path length), we once again find that the only relevant path is the constant path with $\dot{q}(\tau)=0$. Thus, in either the USC limit or the infinite temperature limit, we have,
  \begin{align}\label{eqn:ultra_strong_coupling_mFG_state}
	\rho(q,\eta) = \delta(\eta) Z^{-1} e^{ -\beta V(q) },
\end{align}
where now $Z=\int dq e^{ -\beta V(q) }$. This is the same USC result obtained in \cite{cresser2021weak} and is the CV equivalent of the high temperature result given in \cite{timofeev2022hamiltonian}. Note that the argument used above to obtain the USC-MFGS is equally applicable to discrete systems. See Appendix.~\ref{Appendix:special cases} for more details.

\begin{figure}[htbp]
    \centering
    \includegraphics[width=\linewidth]{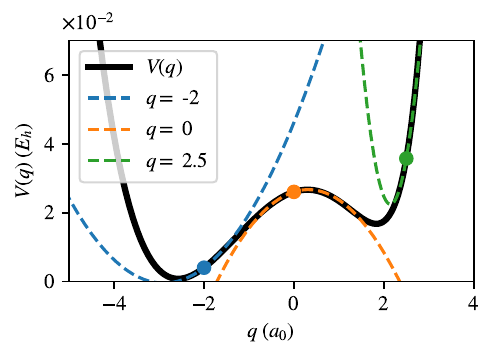}
    \caption{Fitting a harmonic potential at each point $q$ of a general potential $V(q)$.}
    \label{fig:harmonic_fitting}
\end{figure}

Given the discussion above, when either the coupling or temperature is large, but not infinite, we expect that only paths with small deviation around the constant path will contribute significantly to the path integral. In other words, reparameterising the path variable as $q(\tau)=q+\delta(\tau)$, we expect that only small values of $\delta(\tau)$ will be relevant to the final result. Therefore, a local harmonic approximation (LHA) can be employed for $\rho(q,\eta)$ \cite{PhysRevE.81.021107}, where for large coupling or temperature, the effect of the potential locally around $q$ is effectively harmonic. Then, to second order in $\delta(\tau)$, we can write
\begin{multline} \label{eqn:potential expansion}
V(q+\delta(\tau)) \\ \approx  V(q) +\delta(\tau) V_1(q) +\frac{1}{2} \delta(\tau)^2 V_2(q) \\  =V(q) -\frac{1}{2}m\omega_q^2 Q_q^2+ \frac{1}{2}m\omega_q^2\left(\delta(\tau) + Q_q\right)^2
\end{multline}
where
\begin{align}
    \omega_q &= \sqrt{V_2(q)/m} \label{eqn: omega q definition}\\
    Q_q &= \frac{V_1(q)}{V_2(q)} \label{eqn: displacement definition}
\end{align}
are the frequency and the displacement of the effective harmonic potential, respectively, and $V_n(q)$ is the $n$th derivative of $V(q)$. In Fig.~\ref{fig:harmonic_fitting} we show the effective harmonic potentials found for different values of $q$ using a double-well potential for $V(q)$ as an example.

With \autoref{eqn:potential expansion}, the path integral in \autoref{eqn:general path integral} becomes Gaussian and can be computed exactly. This procedure, following Ref. \cite{PhysRevE.81.021107}, yields the LHA expression: 
\begin{equation}\label{eqn: LHA concise version}
    \rho(q,\eta) = Z_R^{-1}\frac{\exp\left(-\beta V(q) \right)}{\exp\left(-\beta \frac{1}{2}m\omega_q^2 Q_{q}^2\right)}\Tilde{Z}_q\Tilde{\rho}_q(Q_q,\eta)
\end{equation}
where $\Tilde{\rho}_{q}(Q_q,\eta)$ is the MFGS for a HO with frequency $\omega_q$, coupled to the same bath as the original system, and $\Tilde{Z}_q$ is the corresponding partition function for the HO-bath system which depends on $V(q)$ only through the effective frequency, $\omega_q$. We highlight that both $\Tilde{\rho}_q$ and $\Tilde{Z}_q$ are straightforward to compute since the HO-bath system is Gaussian and thus exactly solvable \cite{grabert1988quantum, weiss2012quantum}. Although we have expressed $\rho(q,\eta)$ in terms of the Gaussian state of an HO, we stress that \autoref{eqn: LHA concise version} is generally not Gaussian with respect to $q$. This is because the potential, $V(q)$, remains completely arbitrary and all quantities on the right-hand side of \autoref{eqn: LHA concise version} are explicit or implicit functions of this potential.

While the LHA formalism, as expressed in \autoref{eqn: LHA concise version}, is derived under the assumption of either strong coupling or large temperature, we note that it is exact for arbitrary coupling and temperature when the potential is actually harmonic, i.e. when $V(q)=a q +b q^2$ for some $a$ and $b$, no matter how large $\delta(\tau)$ becomes. We therefore expect \autoref{eqn: LHA concise version} to also be accurate even at weak coupling and low temperature, where $\delta(\tau)$ may not be small, as long as the third and higher derivatives of the potential are relatively small. This gives our result great flexibility in its range of validity.

One possible problem with the LHA is that when $V_2(q) < 0$, e.g., near the maxima of the potential $V(q)$, the effective frequency of the HO defined in \autoref{eqn:potential expansion} becomes imaginary, so the HO is inverted \cite{barton1986quantum, subramanyan2021physics, baskoutas1993quantum} (see the orange curve in Fig.~\ref{fig:harmonic_fitting}). For an inverted HO, the potential is unbounded from below as $q\to \pm \infty$, so the MFGS for the HO, $\tilde{\rho}_q$ in \autoref{eqn: LHA concise version}, is unphysical, unnormalisable, and gives divergent observables, e.g. $\braket{\Hat{q}^2}=\infty$.
However, we show in Appendix.~\ref{Appendix: deviation integral} that as long as $ V_2(q) > -V_P$, for some $V_P>0$, the combination $\tilde{Z}_q \tilde{\rho}_q$ can still be finite, where the quantity $\tilde{Z}_q \tilde{\rho}_q$ amounts to an unnormalised path integral which has its important contribution still coming from a finite deviation from the constant path $q(\tau)=q$, as required for LHA to work for general potentials.

This number, $V_P$, is given by the smallest solution of the following equation in $x$,
\begin{equation}\label{eqn:V_2_constraint}
 	\sum_{n=1}^\infty \frac{x}{y_n-x} = \frac{1}{2},
\end{equation}
where
\begin{equation}
  y_n = m\nu_n^2 +  \int _{0}^{\infty} d\omega \frac{J(\omega)}{\omega}  \frac{2 \nu_n ^2}{\omega^2 + \nu_n ^2},
\end{equation}
and $\nu_n = 2 \pi n k_B T$ are the Matsubara frequencies. Thus, provided that $ V_2(q) > -V_P$, the LHA remains finite even for $V_2(q)<0$ and then can be applied to nontrivial potentials, like the one depicted in Fig.~\ref{fig:harmonic_fitting}. We note that in the large coupling or temperature limits we have assumed in deriving the LHA, for $x<y_1$, the $n=1$ term in \autoref{eqn:V_2_constraint} dominates and we obtain the approximate solution $V_P \approx y_1 / 3$. Since $y_n$ diverges with both coupling and temperature, we find the condition $ V_2(q) > -V_P$ is always satisfied in this limit, consistent with LHA becoming exact there.

In order to estimate the error associated with LHA, let us first define two length scales. For the effective quadratic potential approximated at point $q$, let $q_{cl}(\tau)$ denote the classical path starting and ending at the point $q+\eta$ and $q-\eta$. Then, let $\Delta(q,\eta)$ denote the maximum deviation of the classical path from the constant path $q'(\tau) = q$ and let $\mathcal{E}(q)$ denote the typical quantum deviation of all the paths about $q_{cl}(\tau)$ that have significant contribution to the action. An estimate for $\Delta(q,\eta)$ and $\mathcal{E}(q)$ is derived in Appendix.~\ref{Appendix:Length Scale estimation}.

Then, the leading order correction to LHA is given by $\mathcal{T}(q,\eta) \rho(q,\eta)$, where we have,
\begin{align}
	\mathcal{T}(q,\eta) &\equiv -\beta  \frac{V_3(q)}{3!} \left( \Delta(q,\eta)^3 + 3 \mathcal{E}(q)^2 \Delta(q,\eta) \right) \label{eqn:defn of Txy}
\end{align}
The relative error associated with the expectation value of an observable $\Hat{O}$ is given as [Appendix.~\ref{Appendix: error estimation}],
\begin{align}\label{eqn:defn of epsilon of an observable}
    \epsilon_{\Hat{O}} = \frac{1}{\braket{\Hat{O}}}\int dq \int d\eta \; \Hat{O}(q,\eta) \epsilon(q,\eta) |\rho(q,\eta)|
\end{align}
where we have,
\begin{align}\label{eqn: error in population}
    \epsilon(q,\eta) &= |\mathcal{T}(q,\eta)| + \epsilon_T\\
    \epsilon_T &\equiv \int dq \rho(q) |\mathcal{T}(q)| \label{eqn: main paper total diagonal error}
\end{align}

For LHA to be valid for the matrix element $\rho(q,\eta)$, we need $\mathcal{T}(q,\eta) \ll 1$. We also need $\epsilon_T \ll 1$, which translates into,
\begin{equation}\label{eqn:V_3 bound}
    |V_3(q)| << \frac{3!}{\beta \rho(q) \left| \Delta(q)^3 + 3 \mathcal{E}(q)^2 \Delta(q) \right|}
\end{equation}

If, for a system, \autoref{eqn:V_3 bound} is invalid in some region, this does not make LHA totally inapplicable. As an example, if it is valid in two regions of space, A and B, then LHA can still predict the ratio of populations ${\rho_A}/{\rho_B}$ in these two regions.

Finally, for an off-diagonal density matrix element $\rho(q,\eta)$, the value of the classical deviation, $\Delta(q,\eta)$, is at least as large as $\eta$, and hence as the value of $\eta$ is increased, a stronger constraint on $V_3(q)$ is imposed in \autoref{eqn:V_3 bound}. This sets a limit on the length scale on which LHA can accurately capture spatial coherences.

We summarize by commenting that each term in the action in \autoref{eqn:general path integral} gives rise to a regime in which LHA is accurate, i.e., the kinetic, dissipative, and potential terms each give the high temperature, large coupling, and the `slowly varying potential' limit, respectively.

\section{Application}\label{sec:application}

\subsection{Quartic Oscillator} \label{sec:quartic oscillator}

\begin{figure}
\includegraphics{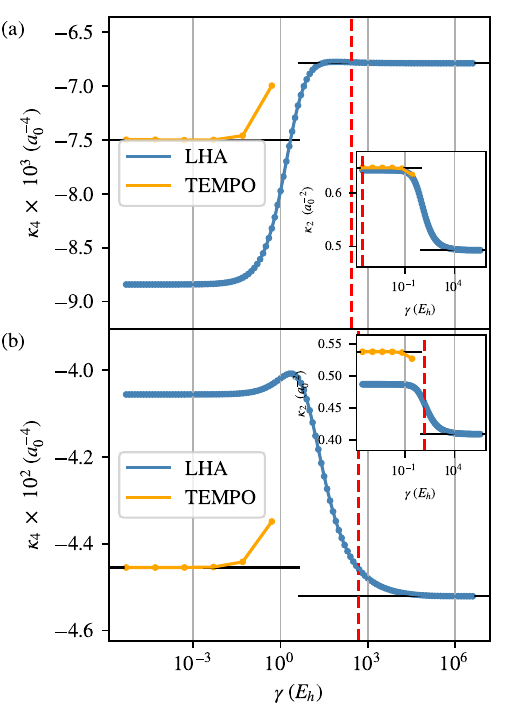}
\caption{The 2nd and 4th cumulant as a function of the system-bath coupling parameter, $\gamma$, for anharmonicity (a) $a=2.5 \times 10^{-3} (E_h/a_0^4)$ and (b) $a = 5 \times 10^{-2} (E_h/a_0^4)$. Orange and blue curves indicate results obtained using TEMPO and LHA, respectively, while the horizontal black lines indicate the exact results for $\gamma=0$ and $\gamma = \infty$. The vertical dashed line corresponds to the point to the right of which $\epsilon_{\kappa_{2/4}} < 0.1$ [\autoref{eqn: kappa_2 error}, \autoref{eqn: kappa4 error}]. Convergence for TEMPO calculations for both (a) and (b) at $\gamma = 0.5 E_h$ was observed for the number of path integral timesteps equal to 60, the system truncation dimension of 9, and the TEMPO SVD precision parameter of $10^{-9}$.}
\label{fig:2}
\end{figure}

We are now going to use LHA to study various CV systems. Let us first study a quartic oscillator whose potential is given by,
\begin{equation}
	V(q) = \frac{1}{2} m \omega^2 q^2  + a q^4\label{eqn:quartic potential}
\end{equation}
A quartic potential is the simplest bounded potential that provides a deviation from the quadratic HO potential, the latter being the case for which LHA works exactly. A quartic oscillator has been studied in the context of quantum to classical transition \cite{Oliveira2006Quantum-classical}, quantum chaos \cite{B.C.Bag1998A, Jt1992Semiclassical, Chakrabarty2019Out} and as a model of bath oscillators interacting with a system \cite{Montenegro2014Nonlinearity}. Here, we study how the non-Gaussianity that naturally emerges in the MFGS of a quartic oscillator changes as the system bath coupling is increased.

We use an Ohmic spectral density of exponential cutoff form (as some integrals required to do the TEMPO calculations done in this section are analytically known in this cutoff),
\begin{align}
	J(\omega) = \frac{2 m \gamma}{\pi} \omega \exp \left( -\frac{\omega}{\Omega} \right)\label{eqn:ohmic spectral density with exp cutoff}
\end{align}
Using Hartree atomic units, we set $\omega = 1 E_h$, $\Omega = 5 E_h$, $m=1 m_e$ and $k_B T = 0.5 E_h$.

To characterize the MFGS, we calculate the second and fourth order cumulants of position, i.e. $\kappa_2=\langle q^2\rangle$ and $\kappa_4 =\langle q^4\rangle-3\langle q^2 \rangle ^2$, respectively. The former of these quantifies the width of the spatial probability density, while the latter quantifies how non-Gaussian the state is, since all cumulants of order three and higher vanish for Gaussian states. The system is symmetric around $q=0$, so the odd-order cumulants of $q$ vanish identically, regardless of whether the state is Gaussian.

To benchmark our results, we use the analytically known weak and USC results, as well as the numerically exact method, TEMPO \cite{strathearn2018efficient, strathearn2020modelling, chiu2022numerical}, applicable in the weak and intermediate coupling regime.

The TEMPO is a tensor network technique to efficiently evaluate the path integral given in \autoref{eqn:general path integral}. It has been previously used to calculate the MFGS for discrete systems \cite{chiu2022numerical}. Here, we use it to study a CV system like a particle in a generic 1D potential $V(q)$. Although the system in question is infinite-dimensional, at any given temperature and coupling, it can be approximated as an effectively finite-dimensional system by truncating the system Hilbert space in the energy eigenbasis \cite{topaler1993system}. The value of this effective dimension can be determined through a convergence test. In this case, for example, for $\gamma = 0.5 \; E_h$, we needed to treat the system as an effectively 9-dimensional discrete system to get sufficient convergence.

\autoref{fig:2} (a) and (b) plot both the cumulants, $\kappa_2$ and $\kappa_4$, calculated using LHA, as a function of coupling strength for $a=2.5 \times 10^{-3}$ and $5 \times 10^{-2}$ [\autoref{eqn:quartic potential}], respectively. The vertical red line marks the value of $\gamma$ such that for $\gamma$ larger than this, the associated relative error $\epsilon_{\kappa_2}, \epsilon_{\kappa_4} < 0.1$, where, from \autoref{eqn:defn of epsilon of an observable} we have,
\begin{align}\label{eqn: kappa_2 error}
    \epsilon_{\kappa_2} &= \epsilon_{\Hat{q}^2}\\
    \epsilon_{\kappa_4} &= \frac{\braket{\Hat{q}^4} \epsilon_{\Hat{q}^4} + 6 \braket{\Hat{q}^2} \epsilon_{\Hat{q}^2}}{\kappa_4} \label{eqn: kappa4 error}
\end{align}

As expected, the accuracy of LHA falls as the value of $\gamma$ is decreased from infinity, while for TEMPO, the computational resources required increase as the value of $\gamma$ is increased from the weak coupling regime.
Because of computational constraints, here we only show the TEMPO results up to $\gamma \leq 0.5$. In \autoref{fig:2} (b), we observe a nontrivial rise in the TEMPO curve for $\kappa_4$ as $\gamma$ is increased, followed by an eventual decay to the USC value as suggested by the LHA curve, suggesting a peak at about $\gamma = 10$.

The results for MFGS in weak and USC limit in \autoref{fig:2} (a) and (b) suggests that the non-gaussianity, that naturally emerges in a quartic oscillator, may either get enhanced or decay as $\gamma$ is increased, depending on the value of the parameter $a$ used in the potential $V(q)$ [\autoref{eqn:quartic potential}]. TEMPO and LHA complement each other to calculate the $\gamma$ dependence of $\kappa_4$ in the low and high $\gamma$ limit, respectively. Note that non-Gaussianity is considered a resource for quantum information processing in CV systems \cite{Takagi2018Convex, weedbrook2012gaussian}, and LHA and TEMPO can help quantify how the non-Gaussianity depends upon the system-bath coupling strength for a CV system.

\subsection{Asymmetric double well potential}\label{sec: double well potential}

Let us now discuss an asymmetric quartic DW potential of the form,
\begin{equation}\label{eqn:quartic double well potential}
    V(q) = a_4 q^4 - a_2 q^2 + a_1 q
\end{equation}
where we set $a_4 = 1/2 = a_2$. In this and the following sections, we use an ohmic spectral density of Drude-Lorentz form,
\begin{align}
	J(\omega) = \frac{2 m \gamma}{\pi} \omega \frac{\Omega^2}{\omega^2 + \Omega^2}\label{eqn:ohmic spectral density with Drude cutoff}
\end{align}
as this form of spectral density admits a closed form expression for the LHA \cite{weiss2012quantum}. We again set $\omega = 1 E_h$, $\Omega = 5 E_h$, $m=1 m_e$ and $k_B T = 0.5 E_h$.

As the accuracy of LHA increases with increasing system bath coupling, we will instead test its applicability in the weak coupling limit. At $\gamma =0$, we are interested in the LHA value of the probability, $\mathcal{P}$, of finding the particle in the right well, given as,
\begin{equation}\label{eqn:quartic right well population definition}
    \mathcal{P} = \int_{q_b}^\infty \rho(q)dq
\end{equation}
where $q_b$ represents the position of the peak of the barrier. We also want to verify the associated estimated relative error for LHA [\autoref{eqn:defn of epsilon of an observable}] against the analytically derivable value for the same at $\gamma =0$.

Since LHA will work at lower coupling only if $V(q)$ varies slowly, let us define,
\begin{equation}\label{eqn: zoomed in potential}
    \Bar{V}_b(q) \equiv b \times V(q/b)
\end{equation}
where for a fixed value of $b>1$, $\Bar{V}_b(q)$ is just a new rescaled potential with smaller values for the higher derivatives with respect to $q$. Note that $\Bar{V}_b(q)$ has been mathematically constructed just to verify the validity of the error estimate [\autoref{eqn:defn of epsilon of an observable}].

\begin{figure}
\includegraphics{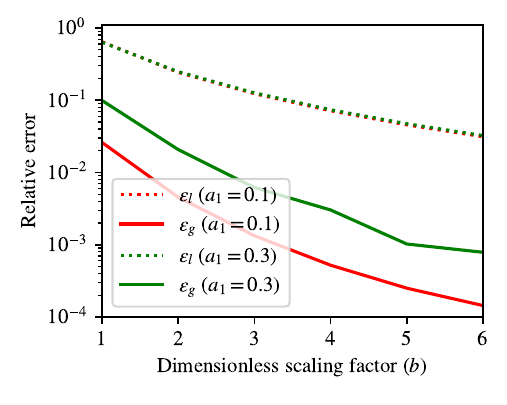}
\caption{Plot for relative error  (solid lines) [\autoref{eqn:actual error quarticDW}] for LHA value of the Gibbs state probability $\mathcal{P}$ [\autoref{eqn:quartic right well population definition}] for the particle to be found in the right well of an asymmetric double well potential $\Bar{V}_b(q)$ [\autoref{eqn: zoomed in potential}], as a function of $b$, a dimensionless parameter of $\Bar{V}_b(q)$. The dotted lines represent the estimated relative error for the same [\autoref{eqn:defn of epsilon of an observable}]. Plots are for $a_1 = 0.1$ (red lines) and $0.3$ (green lines), where $a_1$ quantifies the degree of asymmetry in $\Bar{V}_b(q)$. Note that both dotted lines are almost overlapping. }
\label{fig: zoom in error estimate}
\end{figure}

At a given value of $b$, let $\rho_b$ represent the LHA state and $\sigma_b = Z_b^{-1} \exp\{ -\beta H_0\}$ represent the Gibbs state with $\mathcal{P}_l(b)$ and $\mathcal{P}_g(b)$ being the corresponding values of the probability of finding the particle in the right well, respectively. Here, $Z_b$ is an appropriate normalization factor. Then, let the estimated and analytical value of the relative error be denoted as $\epsilon_l(b)$ [\autoref{eqn:defn of epsilon of an observable}] and $\epsilon_g(b)$, respectively, where $\epsilon_g(b)$ is defined as,
\begin{align}
    \epsilon_g(b) &= \frac{|\mathcal{P}_l(b) - \mathcal{P}_g(b)|}{\mathcal{P}_g(b)}\label{eqn:actual error quarticDW}
\end{align}
In \autoref{fig: zoom in error estimate}, we plot $\epsilon_l(b)$ and $\epsilon_g(b)$ as a function of $b$ for $a_1 = 0.1$ and $0.3$ [\autoref{eqn:quartic double well potential}]. As expected, as the value of $b$ increases and the potential becomes smoother, both $\epsilon_l(b)$ and $\epsilon_g(b)$ decrease in value. We find that $\epsilon_l(b)$ provides an upper bound on $\epsilon_g(b)$, roughly overestimating it by about two orders of magnitude in this case. One possible explanation for this could be that the actual error associated with the expectation value of an observable $\Hat{O}$ can get canceled out during the act of taking the trace, $\Tr{\left( \Hat{O} \rho \right)}$.

\begin{figure}
\includegraphics{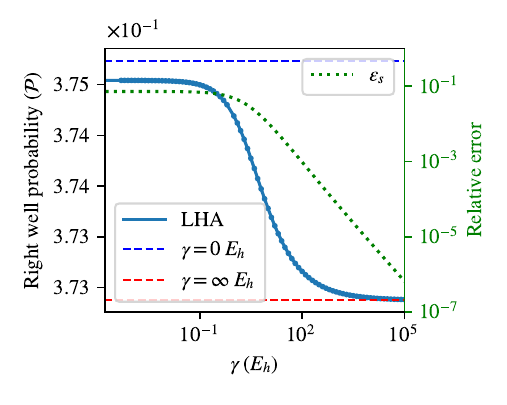}
\caption{The left y-axis plots the LHA value for the probability $\mathcal{P}$ (solid blue line) for a particle in an asymmetric double well potential $\Bar{V}_b(q)$ [\autoref{eqn: zoomed in potential}] to be found in the right well [\autoref{eqn:quartic right well population definition}], as a function of coupling parameter $\gamma$, for $b=4$. The dashed blue and red lines mark the value of $\mathcal{P}$ corresponding to $\gamma = 0$ and $\infty$, respectively. The right y-axis plots the estimated relative error for the corresponding value of $\mathcal{P}$ [\autoref{eqn:defn of epsilon of an observable}].}
\label{fig: mutated population quartic}
\end{figure}

In \autoref{fig: mutated population quartic}, we fix $b=4$ and plot $\mathcal{P}_l(b=4)$ as a function of $\gamma$. We note that throughout the range of $\gamma$, the relative error, $\epsilon_l$, is always less than or equal to $0.07$, indicating that LHA can accurately calculate the value of $\mathcal{P}$ for this DW system throughout the range of the coupling parameter. Note that for this DW system, the values for the distance between the wells, the barrier height from the potential minima, and the energy difference between the well minima are given as $5.6 a_0$, $0.81 E_h$, and $0.56 E_h$, respectively. Hence, although these parameters are not extreme, we note that this DW potential was constructed to be slowly varying with respect to $q$ so that LHA would give accurate results for it, and hence, at best, provides evidence for the accuracy of LHA when $V(q)$ varies slowly.

On the right y-axis of \autoref{fig: mutated population quartic}, we plot the relative error estimate $\epsilon_l(\gamma)$. We note that, as expected, $\epsilon_l(\gamma)$ monotonically decreases with increasing value of $\gamma$. Note that for $\gamma \lesssim 0.1 E_h$, the value of $\epsilon_l(\gamma)$ plateaus, suggesting that in low $\gamma$ limit, the kinetic term in the action [\autoref{eqn:general path integral}] dominates over the dissipative term, and is the sole ($\gamma$ independent) factor that is still responsible for the accuracy of LHA.

\subsection{Proton tunneling in DNA}\label{sec: proton tunneling in DNA}

Slocombe et al. \cite{slocombe2022open, slocombe2022quantum} performed an open quantum system analysis of a proton in Guanine-Cytosine (G-C) pair in DNA to study the probability for it to tunnel to a mutated state. They treated the problem as a particle in an asymmetric double well potential $V(q)$ (see \autoref{fig:harmonic_fitting}), interacting with a non-interacting bosonic bath through an Ohmic spectral density [\autoref{eqn:ohmic spectral density with Drude cutoff}]. Here, $V(q)$ is taken to be a back-to-back Morse potential,
\begin{multline} \label{eqn:SimpleDWpot}
  V(q) = v_1 \left( e^{-2a_1(q-r_1)} - 2e^{-a_1(q-r_1)} \right)  \\ +  v_2 \left( e^{-2a_2(r_2-q)} - 2e^{-a_2(r_2-q)} \right)
\end{multline}
where the lower and higher wells represent the canonical and mutated state for the proton to be in, respectively. Here, in Hartree atomic units, $v_1 = 0.1617 E_h$, $v_2 = 0.082 E_h$, $a_1 = 0.305 a_0^{-1}$, $a_2 = 0.755 a_0^{-1}$, $r_1 = -2.7 a_0$ and $r_2 = 2.1 a_0$. The parameters of the spectral density [\autoref{eqn:ohmic spectral density with Drude cutoff}] were taken to be $\gamma = 0.018 E_h \equiv \gamma_p $ and $k_B T = 0.00095 E_h$ while an infinite bath cutoff was assumed as a Markovian approximation (i.e., $\Omega = \infty E_h$) \cite{slocombe2022quantum}. Here, $\gamma_p$ denotes the biologically relevant value of $\gamma$ for this system.
They calculated the steady state for this system using the Caldeira Leggett master equation (CLME) and found the probability for the proton to tunnel to the mutated state, defined as
\begin{equation}
    \mathcal{M} = \int_{q_b}^\infty \rho(q)dq
\end{equation}
to be of the order of $10^{-4}$, which, as a point of comparison, is several orders of magnitude higher than the zero coupling value of about $2.7 \times 10^{-8}$. Here $q_b$ denotes the position of the barrier maxima.

We remark that CLME is a weak-coupling master equation, which requires $\gamma_p << K_B T, \Omega$. But, for this system, $\gamma_p > K_B T$ and hence the system is well outside the weak coupling regime. This motivates us to study the MFGS of this system using LHA.

\begin{figure}
	\includegraphics{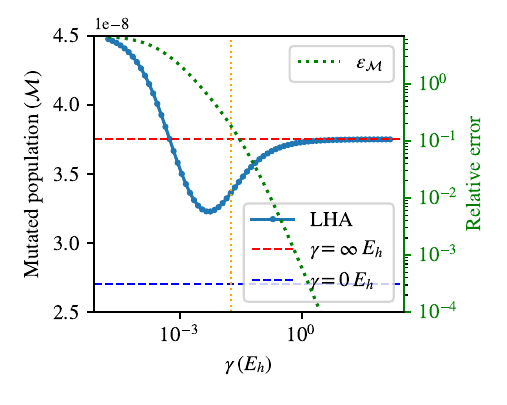}
	\caption{Left y-axis plots the LHA value for mutation probability of the proton, $\mathcal{M}$ (solid blue line), as a function of the coupling parameter $\gamma$. The dashed blue and red lines mark the value of $\mathcal{M}$ corresponding to $\gamma = 0$ and $\infty$, respectively. The vertical orange dotted line marks $\gamma_p = 0.018 E_h$ (the biologically relevant value of $\gamma$). The green dotted curve corresponding to the right-hand side y-axis quantifies the estimated relative error ($\epsilon_{\mathcal{M}}$ [\autoref{eqn: proton relative error}]) in the LHA value for $\mathcal{M}$.}
	\label{fig:mutated_pop_plot_proton_pot}
\end{figure}

In \autoref{fig:mutated_pop_plot_proton_pot}, we plot LHA value for $\mathcal{M}$ as a function of $\gamma$ along with the estimated relative error, $\epsilon_{\mathcal{M}}$ [\autoref{eqn:defn of epsilon of an observable}],
\begin{align}\label{eqn: proton relative error}
    \epsilon_{\mathcal{M}} = \frac{1}{\mathcal{M}}\int_{q_b}^\infty \epsilon(q) \rho(q) dq
\end{align}
We note that at $\gamma = \gamma_p$, $\epsilon_{\mathcal{M}} \approx 0.19$. while for $\gamma \geq 10 \gamma_p$, $\epsilon_{\mathcal{M}} \ll 1$. Hence, in the region where LHA is applicable ($\gamma \geq 10 \gamma_p$), as the value of $\gamma$ is lowered from infinity, we see the value of $\mathcal{M}$ reducing from the USC value, suggesting that $\mathcal{M} = 10^{-4}$ is unlikely. But at $\gamma = \gamma_p$, we can't speculate on the true value of $\mathcal{M}$ since the error is high.

This disagreement could have a couple of possible reasons.
Note that in the intermediate coupling regime, as in this case, it is still unclear whether MFGS is actually the steady state of the system or not \cite{cresser2021weak, trushechkin2022open}. On the other hand, CLME, being a weak-coupling master equation, could be responsible for this disagreement.

\begin{figure}    \includegraphics{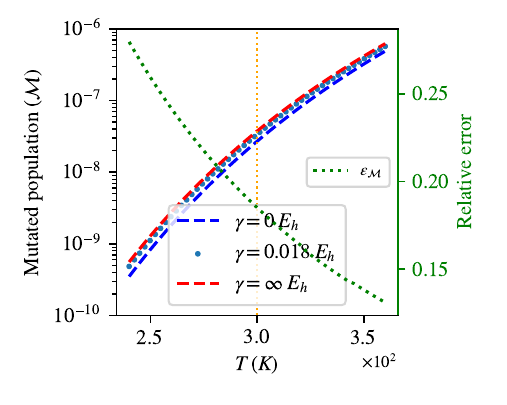}
	\caption{Left y-axis plots the LHA value for the mutation probability, $\mathcal{M}$, as a function of temperature $T$ (in Kelvin) for biologically relevant values of the coupling parameter ($\gamma_p = 0.018 \; E_h$) (blue scatter points). The dashed blue and red lines mark the value of $\mathcal{M}$ corresponding to $\gamma = 0$ and $\infty$, respectively. The vertical orange dotted line marks $T = 300 K$ (i.e., room temperature). The green dotted curve corresponding to the right-hand side y-axis quantifies the estimated relative error ($\epsilon_{\mathcal{M}}$ [\autoref{eqn: proton relative error}]) in the LHA value for $\mathcal{M}$.}
	\label{fig:proton population temperature}
\end{figure}

\autoref{fig:proton population temperature} presents a plot of $\mathcal{M}$ as a function of temperature $T$, across a temperature range of approximately $300 K \pm 60 K$.
As expected, we find that $\mathcal{M}$ rises roughly exponentially with the value of $T$. We also plot the value of $\mathcal{M}$ corresponding to $\gamma = 0$ and $\gamma = \infty$, finding it to closely follow the same trend. At the temperature of about $87^\circ\mathrm{C}$, we find an approximately 10 times rise in the value of $\mathcal{M}$ as compared with the corresponding room temperature value. The accuracy of LHA is observed to increase with a rise in temperature, as expected [Appendix.~\ref{Appendix: High temperature limit}], as observed from the decay in the plot for the relative error $\epsilon_{\mathcal{M}}$ with increasing temperature.

\section{Conclusion}\label{sec: Conclusion}
For a particle in a 1D potential $V(q)$, we first review the Local Harmonic approximation (LHA) to the MFGS [\autoref{eqn:definition of MFGS}].
We derive an estimate for the error induced by this method. We then apply this method to study some systems, like a quartic oscillator and a particle in a quartic double-well potential. We also apply this method to analyze the proton tunneling problem in a DNA recently studied in literature, where our results suggest the equilibrium value of the probability of mutation to be orders of magnitude lower than the steady state value obtained there ($10^{-8}$ vs $10^{-4}$) \cite{slocombe2022open}.
Finally, we have noted that for a CV system, the USC \cite{cresser2021weak} and high temperature \cite{timofeev2022hamiltonian} results obtained recently are a special case of LHA [\autoref{eqn: LHA concise version}]. Finally, we also investigate and confirm the validity of the LHA in the non-intuitive regime of inverted local potentials.

As a future direction, it would be interesting to apply a similar approach to use the analytically known exact expression for the master equation for a harmonic oscillator \cite{hu1992quantum} to derive an approximate master equation for a particle in a generic potential $V(q)$. 
Finally, different ways to improve on LHA and get better error bounds for it will enhance its applicability to physical problems.

\section{Acknowledgement}
I would like to thank my PhD supervisor, Dr. Sibasish Ghosh, for his comments, suggestions, and several extended discussions. I also thank Aidan Strathearn, who has contributed significantly to the preparation of this manuscript.

\bibliographystyle{IEEEtran}
\bibliography{./citation}

\clearpage  

\begin{widetext}

\begin{appendices}
\makeatletter
\renewcommand{\thesection}{\Alph{section}} 
\renewcommand{\thesubsection}{\arabic{subsection}} 
\renewcommand{\thesubsubsection}{\arabic{subsubsection}} 
\makeatother

\section{Summary of the LHA Formalism}\label{Appendix: LHA: Generalization and explicit dependence on $V(q)$}

In this section, we will summarize the basic results of this paper. For details, see the sections mentioned here.

The path integral representation for the MFGS, $\rho(q ,\eta) =\braket{q + \eta |\hat{\rho}|q - \eta}$, is given as,
\begin{equation}\label{eqn:general path integral_a}
	\rho(q ,\eta) = Z_R^{-1} \int_{q + \eta}^{q- \eta} \mathcal{D}q(\tau) e^{-S[q(\tau)]}
\end{equation}
Parameterising the path variable in terms of its deviation from the constant path, $q(\tau) = q + \delta(\tau)$, we expand the potential to second order in $\delta$ [\autoref{eqn:potential expansion}], resulting in \autoref{eqn:general path integral_a} becoming,
\begin{align}\label{eqn:general path integral 2}
	\rho(q ,\eta) &= Z_R^{-1}  \exp\left(-\beta V(q) + \beta \frac{1}{2}m\omega_q^2 Q_q^2\right) \int_{\eta}^{-\eta} \mathcal{D}\delta(\tau)e^{-S_{q}[\delta(\tau) + Q_q]}
\end{align}
where $S_{q}[\delta(\tau) + Q_q]$ is the action for a HO with frequency $\omega_q$ [\autoref{eqn: omega q definition}]. Reparametrize $ \delta(\tau) + Q_q \to \delta(\tau)$ to get,
\begin{align}
    \int_{\eta}^{-\eta} \mathcal{D}\delta(\tau)e^{-S_q[\delta(\tau)+Q_q]} &= \int_{Q_q + \eta}^{Q_q - \eta} \mathcal{D}\delta(\tau)e^{-S_q[\delta(\tau)]}\\
    &= \Tilde{Z}_q \Tilde{\rho}_{q}(Q_q,\eta)
\end{align}
where, $\Tilde{\rho}_{q}(q,\eta)$ is the MFGS for a HO with frequency $\omega_q$. That is, the functional integral in this equation is identical to that of a HO but for the reduced HO partition function $\Tilde{Z}_q$.
This leads us to [\autoref{eqn: LHA concise version}],
\begin{equation}\label{eqn: generalized LHA concise version}
    \rho(q,\eta) = Z_R^{-1}\frac{\exp\left(-\beta V(q) \right)}{\exp\left(-\beta \frac{1}{2}m\omega_q^2 Q_{q}^2\right)}\Tilde{Z}_q\Tilde{\rho}_q(Q_q,\eta)
\end{equation}
We can write the expression for LHA more explicitly as,
\begin{align}
    \rho(q,\eta) &= Z^{-1} \mathcal{G}(q)\exp \left( - \frac{1}{2 \braket{\hat{X}^2(q)}}\left( \frac{V_1 (q)}{V_2(q) } \right)^2  - 2\braket{\hat{P}^2(q)} \eta^2 -\beta \left( V(q)-\frac{V_1 (q)^2}{2 V_2(q)} \right)  \right)\label{eqn:HO_approx}
\end{align}
Here, $\braket{\hat{X}^2(q)}$ and $\braket{\hat{P}^2(q)}$ are the corresponding effective $\braket{\Hat{q}^2}$ and $\braket{\Hat{p}^2}$ values of the harmonic potential approximated at point $q$, whose value is given as \cite{weiss2012quantum, grabert1988quantum},
\begin{align}
    \braket{\hat{X}^2(q)} &=\frac{1}{m \beta} \sum_{n=-\infty}^{+\infty} \frac{1}{\mathcal{A}_n(q)} \label{eqn:q^2 of the fitted oscillator}\\
	\braket{\hat{P}^2(q)} &=\frac{m}{\beta} \sum_{n=-\infty}^{+\infty} \frac{\omega_q^2 + \zeta_n}{\mathcal{A}_n(q)}\label{eqn:p^2 of the fitted oscillator}\\
	\mathcal{A}_n(q) &= \omega_q ^2+\nu_n^2+\zeta_n \label{eqn: mathcal A}\\
    \nu_n &\equiv 2 \pi n k_B T \label{eqn: nu_n definition}\\
    \zeta_n &\equiv \frac{1}{m} \int _{0}^{\infty} d\omega' \frac{J(\omega')}{\omega'}  \frac{2 \nu_n ^2}{\omega'^2 + \nu_n ^2}\label{eqn:HO_zeta}
\end{align}
To arrive at \autoref{eqn:HO_approx}, we did the following replacement in \autoref{eqn: generalized LHA concise version}.
The expression for the reduced partition function, $\Tilde{Z}_q$, is given as [Appendix.~\ref{Appendix:Harmonic oscillator partition function}],
\begin{align}\label{eqn: ZHO definition}
    \Tilde{Z}_q &= \mathcal{I}(q) \sqrt{2 \pi \braket{\hat{X}^2(q)} }
\end{align}
where $\mathcal{I}(q)$ is defined as,
\begin{align}
    \mathcal{I}(q) &\equiv \int_{0}^{0} \mathcal{D}\delta(\tau) e^{-S_q[\delta(\tau)]}\label{eqn:deviation_integral_position}\\
    &= \frac{Z_R}{Z } \mathcal{G}(q) \label{eqn: i q relation}
\end{align}
Here, $Z$ [\autoref{eqn:HO_approx}] is an appropriate constant independent of $q$, and hence is a normalization factor common for all the effective quadratic approximations done at all points $q$. This would imply, for example, that $Z$ can depend upon $k_B T$, $\Omega$, and $\gamma$ but not on $\omega_q$ because the latter explicitly depends upon $V(q)$ by definition [\autoref{eqn: omega q definition}]. The non-trivial part is to then show that $\mathcal{G}(q)$ is given as [Appendix.~\ref{Appendix: deviation integral}],
\begin{align}
    \mathcal{G}(q) &= \frac{1}{\sqrt{ m \omega_q^2 \beta \braket{\hat{X}^2(q)}}}\prod_{n=1}^\infty \frac{\nu_n^2 + \zeta_n }{\mathcal{A}_n(q)} \label{eqn:deviation integral arbitrary potential}
\end{align}

Furthermore, the expression for $\Tilde{\rho}_{q}(q,\eta)$ is given as \cite{weiss2012quantum, grabert1988quantum},
\begin{align}\label{eqn: MFGS for HO}
    \Tilde{\rho}_{q}(q,\eta) &= \frac{1}{\sqrt{2 \pi \braket{\hat{X}^2(q)}}}\exp{\left( - \frac{q^2}{2\braket{\hat{X}^2(q)}} - 2 \braket{\hat{P}^2(q)} \eta ^2 \right)}
\end{align}

With these replacements, we can move from \autoref{eqn: generalized LHA concise version} to \autoref{eqn:HO_approx}. We note that, for a specific spectral density, like Ohmic spectral density with Drude-Lorentz cutoff [\autoref{eqn:ohmic spectral density with Drude cutoff}], the expressions for \autoref{eqn:q^2 of the fitted oscillator}, \autoref{eqn:p^2 of the fitted oscillator}, and \autoref{eqn:deviation integral arbitrary potential} can be evaluated in closed form \cite{weiss2012quantum}.

Note that when $\omega_q^2<0$, $\braket{\hat{X}^2(q)}$ and $\braket{\hat{P}^2(q)}$, as defined in \autoref{eqn:q^2 of the fitted oscillator} and \autoref{eqn:p^2 of the fitted oscillator}, can no more be interpreted as expectation values $\braket{\Hat{q}^2}$ and $\braket{\Hat{p}^2}$ for the corresponding inverted HO, and must instead be taken as mathematical objects as defined above. In Appendix.~\ref{Appendix: deviation integral}, we show that as long as 
\begin{equation} \label{eqn:condition on omegaq}
    m \omega_q^2 > -V_P
\end{equation}
where $V_P$ is defined in \autoref{sec:motivation}, the unnormalized path integral for an inverted HO,
\begin{align}\label{eqn: unnormalized path integral}
    \Tilde{Z}_q \Tilde{\rho}_{q}(Q_q,\eta) = \int_{Q_q + \eta}^{Q_q - \eta} \mathcal{D}\delta(\tau)e^{-S_q[\delta(\tau)]}
\end{align}
is still convergent, which is what we require in the expression of LHA [\autoref{eqn: generalized LHA concise version}]. We note here that $\Tilde{\rho}_{q}(Q_q,\eta)$ and $\Tilde{Z}_q$ are both individually undefined for any inverted HO, regardless of \autoref{eqn:condition on omegaq}.
Given \autoref{eqn:condition on omegaq}, to show that \autoref{eqn: unnormalized path integral} converges, it is sufficient to show the convergence of $\mathcal{G}(q)$, $\braket{\hat{X}^2(q)}^{-1}$ and $\braket{\hat{P}^2(q)}$ that occur in the expression for LHA [\autoref{eqn:HO_approx}].

The expression for the error estimate associated with LHA [\autoref{eqn:defn of epsilon of an observable}] is derived in Appendix.~\ref{Appendix: error estimation}. The expressions for $\Delta(q,\eta)$ and $\mathcal{E}(q)$, required during the error estimation, are given as [Appendix.~\ref{Appendix:Length Scale estimation}],
\begin{align}\label{eqn:classical_deviation_arbitrary_potential}
	\Delta(q,\eta) = \max \left\{ \left |
	\frac{V_1(q)}{V_2(q)} \left ( \frac{W(q)}{\braket{\hat{X}^2(q)}} - 1\right) \right|,
 	| \eta |
	\right\}\\
	\mathcal{E}(q) = \sqrt{ \braket{\hat{X}^2(q)}  - \frac{1 }{V_2(q) \beta}} \left( \sqrt{\frac{1 }{V_2(q) \beta \braket{\hat{X}^2(q)} }} + 1 \right)\label{eqn:quantum_deviation_arbitrary_potential}
\end{align}
where,
\begin{align}   
	W(q) &= \frac{1}{m \beta} \sum_{n=-\infty}^{+\infty} \frac{(-1)^n}{\mathcal{A}_n(q)} \label{eqn:fitted_oscillating_sum}
\end{align}

Finally, LHA in USC, high temperature, and infinite cutoff limit are studied in Appendix.~\ref{Appendix:special cases}.

\section{Expression for LHA}\label{Appendix:Results for MFGS of Harmonic Oscillator}
\subsection{Harmonic oscillator partition function}\label{Appendix:Harmonic oscillator partition function}

In this section, we are going to derive the expression for the HO partition function, $\Tilde{Z}$ [\autoref{eqn: ZHO definition}] for a generic HO potential given as \cite{weiss2012quantum},
\begin{equation}\label{eqn:arbitrary harmonic potential}
	V_{HO}(q) = \frac{1}{2}m \omega^2 q^2
\end{equation}

Let us write any path $q(\tau)$ starting and ending at $q+\eta$ and $q-\eta$, respectively, in terms of the classical path for HO and deviation over it,
\begin{align}\label{eqn:path into classical and deviation position}
	q(\tau) &= q_{\textbf{cl}}(\tau) + \delta(\tau)
\end{align}
where $\delta(0) = 0 = \delta(\beta)$. Then, since the classical path is a local extremum, the action given in \autoref{eqn:general path integral} becomes \cite{feynman1965path},
\begin{align}
	S_{HO}[q(\tau)] &= S_{HO}[q_\textbf{cl}(\tau)] + S_{HO}[\delta(\tau)]\\
	&= S_{HO}[q_\textbf{cl}(\tau)]  + \int_0^\beta d\tau \left( \frac{1}{2} m \dot{\delta}(\tau)^2 + V_{HO}\left(\delta(\tau)\right) \right) + \frac{1}{4}\int_0^\beta d\tau \int_0^\beta d\tau' K\left( \tau - \tau' \right) (\delta(\tau) - \delta(\tau'))^2 \label{eqn:action broken into classical and deviation}
\end{align}

where $S_{HO}[q_\textbf{cl}(\tau)] $ is the action for the corresponding classical path $q_\textbf{cl}(\tau)$ for the HO, given by \cite{weiss2012quantum},
\begin{equation}\label{eqn:classical_action}
	S_{HO}[q_\textbf{cl}(\tau)]  =   \frac{q^2}{2 \braket{\Hat{q}^2}}  + 2 \braket{\Hat{p}^2} \eta^2
\end{equation}

Then, in \autoref{eqn:general path integral}, we have,
\begin{align}\label{eqn:density matrix element position}
	\rho_{HO}(q,\eta) = \Tilde{Z}^{-1} e^{-S_{HO}[q_\textbf{cl}(\tau)]}\mathcal{I}
\end{align}
where, analogous to \autoref{eqn:deviation_integral_position}, we have,
\begin{align}
	\mathcal{I} = \int_0^0 \mathcal{D}\delta(\tau) e^{-S_{HO}[\delta(\tau)]}
\end{align}

Then, using the expression for $\rho_{HO}(q,\eta)$ from \autoref{eqn: MFGS for HO} and using \autoref{eqn:density matrix element position} and \autoref{eqn:classical_action}, we can obtain an expression for $\Tilde{Z}$ as,
\begin{equation}
    \Tilde{Z} = \mathcal{I} \sqrt{2 \pi \braket{\hat{X}^2} }
\end{equation}

\subsection{The path integral for the quantum deviation about the classical path}\label{Appendix: deviation integral}

In this section, we are going to derive the expression for $\mathcal{I}(q)$ [\autoref{eqn:deviation_integral_position}] and consequently $\mathcal{G}(q)$ [\autoref{eqn:deviation integral arbitrary potential}], and also derive the criteria for the convergence of these quantities. Let us do a Fourier series decomposition of $\delta(\tau)$ [\autoref{eqn:path into classical and deviation position}] as \cite{weiss2012quantum, grabert1988quantum},
\begin{align}
	\delta(\tau) &\equiv \frac{1}{\beta}\sum_{n=-\infty}^{\infty} b_n e^{i \nu_n \tau} \label{eqn:delta tau position}
\end{align}

\subsubsection{Action in terms of Fourier components}

The main result of this subsection, given in \autoref{eqn:expression_for_action}, is to express $S_{HO}[\delta(\tau)]$ [\autoref{eqn:action broken into classical and deviation}] in terms of the Fourier components $b_n$ of $\delta(\tau)$ [\autoref{eqn:delta tau position}]. What follows is some routine but tedious math, and one can skip directly to \autoref{eqn:expression_for_action}.

Given \autoref{eqn:delta tau position}, the kinetic part of the action [\autoref{eqn:action broken into classical and deviation}] can be evaluated using,
\begin{equation}\label{eqn:term1}
\int_0^\beta d\tau \dot{\delta} (\tau)^2  = \frac{1}{\beta^2} \int_0^\beta d\tau
 \left ( \frac{d}{d\tau}\sum_{n=-\infty}^{\infty}b_n e^{i \nu_n \tau} \right )^2 = \frac{1}{\beta} \sum_n  \nu_{n}^2 |b_{n}|^2
\end{equation}
Here, we have used the fact that $b_{-n} = b_n ^*$. The potential part of \autoref{eqn:action broken into classical and deviation} can be evaluated as,
\begin{align}
	 \int_0^\beta \delta(\tau)^2 = \frac{1}{\beta^2} \left ( \sum_{n=-\infty}^{\infty}b_n e^{i \nu_n \tau} \right )^2 = \frac{1}{\beta} \sum_n  |b_{n}|^2 \label{eqn:term2}
\end{align}

In order to evaluate the dissipative term, let us first expand the following expression,
\begin{align}
	(\delta(\tau) - \delta(\tau'))^2 &= \frac{1}{\beta^2} \left( \sum_n b_n e^{i \nu_n \tau} - \sum_m b_m e^{i \nu_m \tau'} \right )^2\\
    &= \frac{1}{\beta^2} \sum_p \sum_q b_p b_q e^{i (\nu_p + \nu_q) \tau} \nonumber + \sum_p \sum_q b_p b_q e^{i (\nu_p + \nu_q) \tau'} \nonumber - 2 \sum_p \sum_q b_p b_q e^{i (\nu_p \tau + \nu_q \tau')}
\end{align}

Furthermore, let us first Fourier expand the influence functional as,
\begin{equation} \label{eqn:influence_functional_Fourier_component}
	K(\tau - \tau') = \frac{m}{\beta}\sum_k \xi_k e^{i \nu_k (\tau - \tau')}
\end{equation}
where we have \cite{weiss2012quantum},
\begin{align}\label{eqn:defn of xi}
    \xi_n &= \frac{1}{m} \int_0 ^\infty d\omega J(\omega) \frac{2 \omega}{\nu_n ^2 + \omega^2}
\end{align}

The dissipative term can then be evaluated as,
\begin{align}
    \int_0^\beta d\tau \int_0^\beta d\tau' K\left( \tau - \tau' \right) (\delta(\tau) - \delta(\tau'))^2 = \frac{1}{\beta^2} \int_0^\beta d\tau \int_0^\beta d\tau' \frac{m}{\beta}\sum \xi_k e^{i \nu_k (\tau - \tau')} \left( \sum b_n e^{i \nu_n \tau}  - \sum b_m e^{i \nu_m \tau'}  \right)^2
\end{align}

\begin{align}
	&= \frac{m}{\beta^3} \int_0^\beta d\tau \int_0^\beta d\tau' \sum \xi_k e^{i \nu_k (\tau - \tau')} \left( \sum_p \sum_q \left( b_p b_q e^{i (\nu_p + \nu_q) \tau} + b_p b_q e^{i (\nu_p + \nu_q) \tau'} - 2 b_p b_q e^{i (\nu_p \tau + \nu_q \tau')} \right)\right)\\
    &= \frac{m}{\beta^3} \left( \beta \int_0^\beta d\tau \sum_{p,q} 2 \xi_0 b_p b_q e^{i (\nu_p + \nu_q) \tau}  - 2  \int_0^\beta d\tau \int_0^\beta d\tau' \sum_{p,q,k} \xi_k b_p b_q e^{i (\nu_p+ \nu_k) \tau + i (\nu_q - \nu_k) \tau'} \right)\\
	&= \frac{m}{\beta} \left( \sum_{p} 2 \xi_0 b_p b_{-p} - 2 \sum_{p,q,k} \delta(p+k) \delta(q-k)\xi_k b_p b_q \right)\\
	&=\frac{2m}{\beta} \sum_{p}  (\xi_0 - \xi_p) |b_p|^2 \quad \text{sum over $p$ and $q$ then rename $k \to p$}\\
	&=\frac{2m}{\beta} \sum_{p}  (\zeta_p) |b_p|^2 \label{eqn:term3}
\end{align}
where we used $\zeta_p = \xi_0 - \xi_p$ [\autoref{eqn:HO_zeta}], which can be derived from \autoref{eqn:defn of xi} as,
\begin{align}
	\xi_0 -\xi_n &= \frac{1}{m} \int_0 ^\infty d\omega \frac{J(\omega)}{\omega} \left( \frac{2 \nu_n ^2 }{ (\nu_n ^2 + \omega^2)} \right) = \zeta_n \qquad \text{from \autoref{eqn:HO_zeta}}
\end{align}

Finally, using the simplified expressions for the kinetic [\autoref{eqn:term1}], potential [\autoref{eqn:term2}] and dissipative [\autoref{eqn:term3}] part of the action \autoref{eqn:action broken into classical and deviation}, we have,
\begin{align}
	S_{HO}[\delta(\tau)] &= \sum_{n = -\infty} ^\infty \frac{1}{2} \frac{m}{\beta} |b_n|^2 \left ( \nu_n ^2 + \omega^2 + \zeta_n\right ) \label{eqn:expression_for_action}
\end{align}

\subsubsection{Calculating the deviation integral}\label{Appendix:Deviation when the second derivative of the potential is negative}

Let us define,
\begin{align}
	\alpha_n &\equiv \frac{m}{\beta} ( \omega^2 + \nu_n ^2 +  \zeta_n)\label{eqn:expression_for_alpha_n}
 \end{align}
The factor $\frac{1}{\alpha_n}$ gives an effective cutoff for the value of $|b_n^2|$ for each value of $n$. For any $n$, if $|b_n|^2$ crosses this cutoff, then the influence of the corresponding path will decay rapidly. Let us move to the Cartesian coordinates, defining $x_n$ and $y_n$ as the real and imaginary parts of the complex number $b_n$ (i.e.,
$x_n \equiv \Re (b_n) $ and
$y_n \equiv \Im(b_n)$). Since $\delta(0) = 0$, the Fourier component of any function of this type must satisfy,
\begin{align} \label{eqn: b_n constraint}
	\sum_{n=-\infty}^{\infty}x_n &= 0\\
	\implies x_0 &= -2  \sum_{n =1}^\infty x_n \label{eqn:constraint_on_b_n_positive}
\end{align}
Hence, $x_0$ is not an independent parameter in the path integral. Similarly, since $x_{-n} = x_{n}$ and $y_{-n} = -y_{n}$, hence the only independent variables are those corresponding to $n\geq1$. The path integral $\mathcal{I}$ [\autoref{eqn:deviation_integral_position}] then becomes,
\begin{align}\label{eqn: deviation integral cartesian coordinate}
	\mathcal{I} &= J \left( \prod_{n = 1}^\infty \int_{-\infty}^\infty  d y_n \int_{-\infty}^\infty d x_n \right) \exp \left\{ -\frac{1}{2} \alpha_0 \left( 2 \sum_{n=1}^\infty x_n \right)^2 - 2 \sum_{n=1}^\infty \frac{1}{2} \alpha_n \left( x_n^2 + y_n^2 \right) \right\}
\end{align}
Here, $J$ is the Jacobian coming from the linear transformation given in \autoref{eqn:delta tau position}.
We will now analytically evaluate this integral. But we should first note that this integral will diverge if for $n\geq 1$, $\alpha_n \leq 0$, which would break down LHA.

Before proceeding, note that we have
$\mathcal{\alpha}_{|n+1|} > \mathcal{\alpha}_{|n|}$.
This is true because $\nu_{|n+1|} > \nu_{|n|}$ follows from definition and immediately gives $\zeta_{|n+1|} > \zeta_{|n|}$ since $x/(a+x)$ monotonically increases for positive $x$ and $a$. Hence, as a necessary condition for \autoref{eqn: deviation integral cartesian coordinate} to be convergent, we just need to impose that $\alpha_1 >0$. We will hence restrict ourselves to this constraint. (As a side remark, note that this assumption automatically ensures the convergence of $\braket{\hat{P}^2(q)}$ [\autoref{eqn:p^2 of the fitted oscillator}].)
But it is still unclear whether $\alpha_0 \leq 0$ will lead to divergence in \autoref{eqn: deviation integral cartesian coordinate} or not.
In this section, we will see that $\mathcal{I}$ can be convergent even if $\alpha_0 \leq 0$.

After doing the $y_n$ integral in \autoref{eqn: deviation integral cartesian coordinate}, we get,
\begin{align}\label{eqn:cartesian_integral_x_n}
	\mathcal{I} = J \left( \prod_{m=1}^\infty \sqrt{\frac{\pi}{\alpha_m}}\right) \left( \prod_{n =1 } \int_{-\infty}^\infty d x_n \right) \exp\left\{-\left(  2 \alpha_0 \left[ \sum_{n=1}^\infty x_n\right]^2 + \sum_{n=1}^\infty  \alpha_n x_n^2 \right)\right\}
\end{align}

Let us define the negative of the exponent above as,
\begin{align}
  	\mathcal{F} \equiv \left( 2 \alpha_0 \left[  \sum_{n=1}^\infty x_n\right]^2 + \sum_{n=1}^\infty \alpha_n x_n^2 \right)
\end{align}
If $\min_{x_n} \mathcal{F} > 0$, then $\mathcal{I}$ will converge. We can think of $\mathcal{F}$ as a diagonal component of a matrix $M$, i.e.
$\mathcal{F} \equiv \langle v | M | v\rangle$
where,
$\langle v | = \langle x_1, x_2, x_3, ... |$ and
$M  = D + 2 \alpha_0 J$
where, for $i,j \geq 1$, $J_{ij} = 1$ and $D_{ij} = \delta_{ij} \alpha_i$. Let $\lambda_n$ be the eigenvalues of the matrix $M$. Then we have,
\begin{align}
	\mathcal{I} = J \left( \prod_{n=1}^\infty \frac{\pi}{\sqrt{\alpha_n \lambda_n}} \right)
\end{align}

Before proceeding, let us first simplify an expression we will need immediately. From the definition of $\braket{\Hat{q}^2}$ [\autoref{eqn:q^2 of the fitted oscillator}] and $\alpha_n$ [\autoref{eqn:expression_for_alpha_n}], we have,
\begin{align}
	\braket{\Hat{q}^2} &= \frac{1}{\beta^2}\sum_{n=-\infty}^{\infty} \frac{1}{\alpha_n}\\
	\implies 2 \sum_{n=1}^\infty \frac{1}{\alpha_n}  &= \beta^2\braket{\Hat{q}^2} - \frac{1}{\alpha_0}\label{eqn:sum_as_q2}
\end{align}

Proceeding with the derivation of $\mathcal{I}$ again, we next need to calculate the value of the product $\prod_{n=1}^\infty \lambda_n$. We note that this is just given by the determinant of the matrix $M$, i.e.,
\begin{align}
	\prod_{n=1}^\infty \lambda_n &= |M|\\
	&= \left( \prod_{n=1}^\infty \alpha_n \right) + 2 \alpha_0 \left( \prod_{n=1}^\infty \alpha_n \right) \left( \sum_{m=1}^\infty \frac{1}{\alpha_m}\right)\\
	&=  \left( \prod_{n=1}^\infty \alpha_n \right) \left( 1 + \alpha_0 \left( \beta^2\braket{\Hat{q}^2} - \frac{1}{\alpha_0} \right) \right) \qquad \text{from \autoref{eqn:sum_as_q2}}\\
	&= \alpha_0  \beta^2\braket{\Hat{q}^2} \prod_{n=1}^\infty \alpha_n
\end{align}
We hence have,
\begin{align}
    \mathcal{I} &= J \left( \frac{1}{\sqrt{\alpha_0 \beta^2 \braket{\Hat{q}^2}}} \prod_{n=1}^\infty \frac{\pi}{\alpha_n } \right)
\end{align}
This infinite product will give an infinitesimally small result because $\alpha_n$ scales as $n^2$. But to get a reasonable result for LHA, we can proceed as follows. In the expression for LHA [\autoref{eqn: generalized LHA concise version}], we effectively have a factor of $\frac{\mathcal{I}(q)}{Z_R}$. Let us transform it as,
\begin{align}
	\frac{\mathcal{I}(q)}{Z_R} \equiv \frac{\mathcal{G}(q)}{Z}
\end{align}
where we define $Z$ as,
\begin{align}
    Z = Z_R \left( \frac{1}{J}\prod_{n=1}^\infty \frac{\frac{m}{\beta} ( \nu_n ^2 +  \zeta_n) }{\pi}\right)
\end{align}
We hence have, for $\mathcal{G}(q)$,

\begin{align}
    \mathcal{G} &= \frac{Z}{Z_R} \mathcal{I} = \left( \frac{1}{J}\prod_{n=1}^\infty \frac{\frac{m}{\beta} ( \nu_n ^2 +  \zeta_n) }{\pi}\right) \mathcal{I}\\
    &= \frac{1}{\sqrt{\alpha_0 \beta^2 \braket{\Hat{q}^2}}}\prod_{n=1}^\infty \frac{\alpha_n - \frac{m \omega^2}{\beta} }{\alpha_n } \label{eqn:deviation integral solution}
\end{align}
This quantity can be shown to converge. On the other hand, $Z$ serves as an overall normalization factor for LHA since, as we remarked in Appendix.~\ref{Appendix: LHA: Generalization and explicit dependence on $V(q)$}, it depends only upon $k_B T$, $\Omega$ and $\gamma$, but not on $\omega$, and is hence independent of $q$. Note that the Jacobian $J$, coming from the linear transformation given in \autoref{eqn:delta tau position}, is also $\omega$ independent.

Now, we are going to investigate the condition for \autoref{eqn:deviation integral solution} to converge. Since we have assumed $\alpha_{n\geq 1} >0$, we note that for \autoref{eqn:deviation integral solution} to be real, we need,
\begin{align}
    \alpha_0 \braket{\Hat{q}^2} &> 0 \label{eqn:condition on q2}\\
    \implies \alpha_0 \left( \frac{1}{\alpha_0} + 2 \sum_{n=1}^\infty \frac{1}{\alpha_n} \right) &> 0 \qquad \text{from \autoref{eqn:sum_as_q2}}\label{eqn:condition for convergence}
\end{align}
If $\alpha_0$ is also positive, then \autoref{eqn:condition for convergence} is automatically satisfied. If $\alpha_0 < 0$, then we are interested in the highest value of $\alpha_0$ such that we have,
\begin{align}
    \left( \frac{1}{\alpha_0} + 2 \sum_{n=1}^\infty \frac{1}{\alpha_n} \right) &= 0\\
    \implies \sum_{n=1}^\infty \frac{\alpha_0}{\alpha_n} &= -\frac{1}{2}  \label{eqn:condition_on_alpha_0}\\
	\implies \sum_{n=1}^\infty \frac{m \omega^2}{m \omega^2 + m \nu_n + m \zeta_n} &= -\frac{1}{2}
\end{align}
Here, we used the definition of $\alpha_n$ [\autoref{eqn:expression_for_alpha_n}]. Redefining $m \omega^2 = -x$ and $m \nu_n^2 + m \zeta_n = y_n$, we get \autoref{eqn:V_2_constraint}, where now we are interested in the smallest solution for $x$ in the following equation,
\begin{align}\label{eqn:smallest solution for x}
	\implies \sum_{n=1}^\infty \frac{x}{y_n - x} &= \frac{1}{2}
\end{align}
The smallest solution for $x$ gives us the value of $V_P$, defined in \autoref{sec:motivation}, such that LHA will give convergent results for $V_2(q)>-V_P$.

Note that, from \autoref{eqn:condition on q2}, for $\alpha_0<0$, we are equivalently interested in the highest negative value of $\omega^2$ such that $\braket{\Hat{q}^2}=0$. Hence, for the value of $\omega^2$ higher than this, convergence of $\braket{\hat{X}^2(q)}^{-1}$ [\autoref{eqn:q^2 of the fitted oscillator}] is also guaranteed, as required by LHA [\autoref{eqn:HO_approx}] to be well defined. Also, since some spectral densities admit closed form expression for $\braket{\Hat{q}^2}$, this condition can easily be checked for in these cases.

In summary, we can say that LHA gives convergent results if
\begin{itemize}
	\item $\alpha_0 > 0$,
	\item $\alpha_0 < 0$ and $\alpha_1 > 0$, and \autoref{eqn:condition for convergence} is true. The value of $V_P$ is given as the smallest solution for $x$ in \autoref{eqn:smallest solution for x}.
\end{itemize}

\section{Error estimation}\label{Appendix: error estimation}
In this section, we are going to derive the expression for the estimation of the error associated with LHA [\autoref{eqn:defn of epsilon of an observable}].
We start with \autoref{eqn:general path integral}. Parameterising the path variable as $q(\tau)=q+\delta(\tau)$, where $q(\tau)$ starts and ends at points $q+\eta$ and $q-\eta$, respectively, we can write,
\begin{equation}
    V(q+\delta)=V(q) + V_1(q)\delta +\frac{1}{2!}V_2(q)\delta^2 +\frac{1}{3!}V_3(q)\delta^3 +O(\delta^4)
\end{equation}
Let us write the full action in terms of the quadratic part and higher-order terms in the potential, as,
\begin{align}
	S[q + \delta(\tau)] = S_q[q + \delta(\tau)] + \int_0^\beta d\tau \sum_{n=3} \frac{1}{n!}V_n(q)\delta(\tau)^n
\end{align}
Then, writing the exact state $\sigma(q,\eta)$ in terms of harmonic action $S_q[q + \delta(\tau)]$ and higher order corrections over it, we get,
\begin{align}\label{eqn: first order correction to LHA}
    \sigma(q,\eta)= Z'^{-1}\left \{ \int_{ - \eta}^{\eta} \mathcal{D}\delta(\tau) \exp \left(-\int_0^\beta d\tau \sum_{n=3} \frac{1}{n!}V_n(q)\delta(\tau)^n\right)  e^{-S_q[q + \delta(\tau)]} \right \}
\end{align}
Here, $Z'$ is the normalization factor defined as,
\begin{align}
    Z' = \int_{-\infty}^{\infty} dq\left \{ \int_{0}^{0} \mathcal{D}\delta(\tau) \exp \left(-\int_0^\beta d\tau \sum_{n=3} \frac{1}{n!}V_n(q)\delta(\tau)^n\right)  e^{-S_q[q + \delta(\tau)]} \right \}
\end{align}
We define
\begin{align}
    Z &\equiv \int_{-\infty}^{\infty} dq\left \{ \int_{0}^{0} \mathcal{D}\delta(\tau)  e^{-S_q[q + \delta(\tau)]} \right \} 
\end{align}
Next, let us approximate \autoref{eqn: first order correction to LHA} as,
\begin{align}
    \sigma(q,\eta) &\approx \frac{Z}{Z'} \rho(q,\eta) - \frac{1}{Z'}  \sum_{n=3} \frac{V_n(q)}{n!} \left \{ \int_{ - \eta}^{\eta} \mathcal{D}\delta(\tau)  \left(\int_0^\beta d\tau \delta(\tau)^n\right)  e^{-S_q[q + \delta(\tau)]} \right \}\\
    &\approx \frac{Z}{Z'}  \rho(q,\eta) - \frac{1}{Z'}  \frac{V_3(q)}{3!} \left \{ \int_{ - \eta}^{\eta} \mathcal{D}\delta(\tau)  \left(\int_0^\beta d\tau \delta(\tau)^3\right)  e^{-S_q[q + \delta(\tau)]} \right \}
\end{align}
Here, $\rho(q,\eta)$ is the LHA state, and in the last step, we have made a first-order approximation. We now need to calculate the quantity,
\begin{align}\label{eqn: the complicated path integral}
	 I_3 = \int_{- \eta}^{\eta} \mathcal{D}\delta(\tau) \left( \int_0^\beta d\tau \delta(\tau)^3 \right)  e^{-S_q[q + \delta(\tau)]}
\end{align}
Here, we are going to estimate this path integral using the following argument. Reparametrize $q  + \delta(\tau) \to q_{\textbf{cl}}(\tau) + \delta(\tau)$. Then using $S_q[q + \delta(\tau)] \to S_q[q_{\textbf{cl}}(\tau) + \delta(\tau)] = S_q[q_{\textbf{cl}}(\tau)] + S_q[\delta(\tau)]$, we get,
\begin{align}
    I_3 &= e^{-S_q[q_\textbf{cl}(\tau)] } \int_{0}^{0} \mathcal{D}\delta(\tau) \left( \int_0^\beta d\tau \left(q_{\textbf{cl}}(\tau) - q + \delta(\tau)\right)^3 \right)  e^{-S_q[\delta(\tau)]}
\end{align}
Now, if $n$ is an odd number, then we have $(-\delta(\tau))^n = -\delta(\tau)^n$. But since $S_q[-\delta(\tau)] = S_q[\delta(\tau)]$, when we take the full path integral, we get,
\begin{align}
    \int_{0}^{0} \mathcal{D}\delta(\tau) \left( \int_0^\beta d\tau (q_{\textbf{cl}}(\tau) - q)^m \delta(\tau)^n \right)  e^{-S_q[\delta(\tau)]} &= 0 \qquad \text{for odd} \; n
\end{align}
\begin{align}
    \implies I_3 &= e^{-S_{q}[q_\textbf{cl}(\tau)] } \int_{0}^{0} \mathcal{D}\delta(\tau) \left( \int_0^\beta d\tau \left((q_{\textbf{cl}}(\tau)-q)^3 + 3 (q_{\textbf{cl}}(\tau)-q) \delta(\tau)^2\right) \right)  e^{-S_q[\delta(\tau)]}
\end{align}
Now, we are going to replace $q_{\textbf{cl}}(\tau)-q$ with its maximum value, given by $\Delta(q,\eta)$, and $\delta(\tau)$ with its typical value, given by $\mathcal{E}(q)$, to obtain,
\begin{align}
	 I_3 \approx \beta \left( \Delta(q,\eta)^3 + 3  \mathcal{E}(q)^2 \Delta(q,\eta) \right) \int_{- \eta}^{\eta} \mathcal{D}\delta(\tau) e^{-S_q[q + \delta(\tau)]}
\end{align}
We will estimate $\Delta(q,\eta)$ and $\mathcal{E}(q)$ in \autoref{Appendix:Length Scale estimation}.

Let's define a quantity, $\mathcal{T}(q,\eta)$, as a measure of the relative error corresponding to the matrix element $\rho(q,\eta)$ as,
\begin{align}\label{eqn: definition of relative error}
    \mathcal{T}(q,\eta) &\equiv -\beta  \frac{V_3}{3!} \left( \Delta(q,\eta)^3 + 3 \mathcal{E}(q)^2 \Delta(q,\eta) \right)
\end{align}

Also, for convenience, let us define,
\begin{align}
    \epsilon_T &\equiv \int dq \rho(q) |\mathcal{T}(q)|\\
    &= \braket{|\mathcal{T}(q)|}
\end{align}

Then, \autoref{eqn: first order correction to LHA} becomes,
\begin{align}
    \sigma(q,\eta) &\approx \frac{\rho(q,\eta) \pm \rho(q,\eta)  |\mathcal{T}(q,\eta)|}{Z'}\\
    &= \frac{\rho(q,\eta) \pm \rho(q,\eta)  |\mathcal{T}(q,\eta)|}{1 \pm \epsilon_T}\\
    &\approx \rho(q,\eta) \pm \epsilon(q,\eta) \rho(q,\eta)
\end{align}
where,
\begin{align}
    \epsilon(q,\eta) = |\mathcal{T}(q,\eta)| + \epsilon_T
\end{align}
gives a measure of the total relative error for the matrix element $\rho(q,\eta)$. To get an expression for the relative error associated with an observable $\Hat{O}$, we define,
\begin{align}
    \epsilon_{\Hat{O}} = \frac{1}{\braket{\Hat{O}}}\int dq \int d\eta \; \Hat{O}(q,\eta) \epsilon(q,\eta) |\rho(q,\eta)|
\end{align}

\subsection{Length Scale estimation}\label{Appendix:Length Scale estimation}

In the following subsections, we will estimate $\Delta(q,\eta)$ and $\mathcal{E}(q)$.
Let $q_{\textbf{cl}}(\tau)$ be the classical path for an HO, starting and ending at points $q+\eta$ and $q-\eta$, respectively, given by,
\begin{equation}\label{eqn:classical path Fourier expansion}
	q_{\textbf{cl}}(\tau) = \frac{1}{\beta}\sum_{n=-\infty}^\infty a_n e^{i \nu_n \tau}
\end{equation}

Then, $a_n$ is given as \cite{weiss2012quantum},
\begin{align}
	a_n &= \left\{ \left(\frac{i \nu_n}{\omega^2 + \nu_n^2 +  \zeta_n} \right) 2 \eta + \frac{1}{\omega^2 + \nu_n^2 +  \zeta_n} \frac{q}{m \braket{\Hat{q}^2}}  \right\} \label{eqn:classical_path_Fourier_component}
\end{align}

We conjecture that for $\eta = 0$, $|q_{\textbf{cl}}(\tau) - q|$ will have its global maxima at $\tau = \beta/2$. This intuition rests on the fact that in the general path integral equation [\autoref{eqn:general path integral}],  although the potential term will push the extremum classical path towards the potential minima, the kinetic and dissipative terms will help remove any oscillations from it. Hence, the curve will have a single extrema, and since for $\eta = 0$, the classical path can be seen to be made up of only cosine Fourier components and hence is symmetric about $\tau = \beta/2$, therefore this extrema will be at $\tau = \beta/2$. We have also verified this numerically. When $\eta \neq 0$, we note that although the classical path still does not have any oscillations, $\tau = \beta/2$ does not give the extremum point anymore, but we assume that the maximum among $| q_{\textbf{cl}}(\beta/2)-q |$ and $| \eta |$ still gives a rough estimate for the maximum deviation of the classical path from the constant path $q'(\tau) = q$.

We will hence use the quantity $ q_{\textbf{cl}}(\beta/2) - q$ to estimate $\Delta(q,\eta)$. We simplify this expression as,
\begin{align}
	q_{\textbf{cl}}(\beta/2) - q  &=- q + \frac{1}{\beta} \sum_{n=-\infty}^\infty (-1)^n a_n\\
	&= q \left( \frac{1}{m \beta \braket{\Hat{q}^2}} \sum_{n=-\infty}^\infty \frac{(-1)^n}{\nu_n^2 + \omega^2 + \zeta_n}- 1 \right)
\end{align}

Hence, the estimate for the maximum deviation of the classical path is then given by,
\begin{align}
	\Delta_{HO}(q,\eta) = \mathop{\mathrm{max}}\left\{ \left| q \left( - 1 + \frac{1}{m \beta \braket{\Hat{q}^2}} \sum_{n=-\infty}^{\infty} \frac{(-1)^n}{\nu_n^2 + \omega^2 + \zeta_n}  \right) \right| ,  | \eta |\right \} \label{eqn:HO_max_dev_classical_path}
\end{align}
Relating $q \equiv V_1(q)/V_2(q)$ [\autoref{eqn: displacement definition}], we directly recover the expression for the maximum deviation of the classical path claimed earlier [\autoref{eqn:classical_deviation_arbitrary_potential}].

Next, in order to estimate $\mathcal{E}(q)$, rename $r_n \equiv |b_n|$ [\autoref{eqn:expression_for_action}] and write the path integral for $\mathcal{I}$ [\autoref{eqn: deviation integral cartesian coordinate}] in polar coordinate as,
\begin{align}
	\mathcal{I} = J \left( \prod_{n = 1}  \int_{0}^{\infty} d r_n \int_{0}^{2 \pi} r_n d \theta_n \right) e^{-\left( \frac{1}{2}\alpha_0 b_0^2+ 2 \sum_{n=1}^\infty \frac{1}{2} \alpha_n r_n^2 \right)} \label{eqn:Gaussian_integral_form}
\end{align}
Note that $\theta_n$ dependence in the integrand is in $b_0$ and comes through the following constraint [\autoref{eqn: b_n constraint}],
\begin{align}\label{eqn: quantum deviation b_n constraint}
    b_0 = -2 \sum_{n=1}^\infty \Re(b_n)
\end{align}
We are now in the position to immediately do all the $d \theta_n$ integrals in the \autoref{eqn:Gaussian_integral_form}. For this, we note that for a fixed value of $b_n$'s, where $n = 0, 1, 2, ... $, the $d \theta_n$ integrals in \autoref{eqn:Gaussian_integral_form} give us a measure of all the configurations that satisfy the constraint in \autoref{eqn: quantum deviation b_n constraint}.

Now consider a 2D random walk starting from the origin, where the nth step size for the walk is given by the function $\mathcal{S}(n) = 2 |b_n| = 2 r_n$.  Then, after $N \to \infty$ steps, the probability distribution $P(x)$ for this walker to have the x-axis position value of $-b_0$ gives a measure that is equivalent to doing all the $d \theta_n$ integrals in \autoref{eqn:Gaussian_integral_form}. Note that there is a factor of 2 in the definition for $\mathcal{S}(n)$ because of the corresponding factor of 2 in \autoref{eqn: quantum deviation b_n constraint}.

This probability distribution is given as  \cite{sherman1972summation},
\begin{equation}
	P(x) = \frac{1}{\sigma}\sqrt{\frac{1}{2 \pi }} e^{-\frac{x^2}{2 \sigma^2}}
\end{equation}
where, the variance $\sigma^2$ is given by,
\begin{align}
	\sigma^2 &= \frac{1}{2} \sum_{n=1}^\infty (\mathcal{S}(n))^2\\
	&= 2 \sum_{n=1}^\infty r_n^2 \label{eqn:sigma squared}
\end{align}
Note that the Gaussian probability distribution for the random walk is true only if a few sets of steps, $\mathcal{S}(n)$, do not dominate over the others in size. If this is not the case, the actual distribution is more complicated (although it is still a decaying function with some width), and is given in terms of Bessel functions \cite{sherman1972summation}. Here, for simplicity, we assume a Gaussian probability distribution.

We hence replace all the $d_{\theta_n}$ integrals with a single integral over the variable $b_0$, and  \autoref{eqn:Gaussian_integral_form} becomes,
\begin{align}
    \mathcal{I} &= J \left(\prod_{n = 1}^\infty  \int_{0}^{\infty} r_n d r_n \right) \int_{-R}^{R}d b_0  \frac{1}{\sigma}\sqrt{\frac{1}{2 \pi }}  e^{-\left( \frac{1}{2}  \left(\alpha_0 + \frac{1}{\sigma^2}\right) b_0^2 + 2 \sum_{n=1}^\infty \frac{1}{2} \alpha_n r_n^2 \right)} \label{eqn:the_full_integral}
\end{align}
where
$R=2\sum_{n=1}^\infty r_n$
is put here to ensure that \autoref{eqn: quantum deviation b_n constraint} is not violated.

Note that this equation gives a typical length-scale for the value of $r_n$ and $b_0$ as follows,
\begin{align}
	r_n^2 = |b_n|^2 &\approx \frac{1}{\alpha_n}\label{eqn:upper_bound_b_n} \\
	\implies \sigma^2 &\approx 2 \sum_{n=1}^\infty  \frac{1}{\alpha_n} \qquad \text{from \autoref{eqn:sigma squared}} \label{eqn: typical sigma2}\\
	\implies b_0^2 &\approx \frac{1}{\alpha_0 + \frac{1}{2 \sum_{n=1}^\infty  \frac{1}{\alpha_n}}}\label{eqn:upper_limit_on_b_0}
\end{align}

Using the typical value of $|b_n|$ from here, we can go back in the position space representation of $\delta(\tau)$ [\autoref{eqn:delta tau position}] that corresponds to the paths that have an important contribution to $\mathcal{I}$, as,
\begin{align}\label{eqn:delta tau for important paths}
	\delta(\tau) = \frac{1}{\beta} \left( \frac{1}{\sqrt{\alpha_0 + \frac{1}{2 \sum_{n=1}^\infty  \frac{1}{\alpha_n}}}} + \sum_{n\neq0}\frac{1}{\sqrt{\alpha_n}} e^{i \theta_n} e^{i \nu_n \tau} \right)
\end{align}
where, $\theta_{-n} = -\theta_n$ are random phases with the condition, coming from \autoref{eqn: quantum deviation b_n constraint}, that $\delta(\tau=0) = 0$.

We are interested in the typical deviation of $\delta(\tau)$. To simplify the problem, we now relax the condition $\delta(\tau=0) = 0$ so that $\theta_n$ becomes truly random,m and this again reduces to the random walk problem with fixed step size \cite{sherman1972summation}. To be precise, the mathematical problem we are now dealing with is- given a Fourier series with fixed coefficients but random phases $\phi_n$,
\begin{align}
	h(x) = c_0 + \sum_{n=1}^\infty c_n \sin{\left( 2 \pi n x + \phi_n \right)}
\end{align}
What is the probability density, $P(y)$, that $h(x) = y$ for any given value of $x$?

Using similar random walk arguments as used before, $P(y)$ can be shown to be approximately given by,
\begin{align}
	P(y) = \frac{1}{\sigma}\sqrt{\frac{1}{2 \pi}} e^{\frac{-(y-c_0)^2}{2 \sigma^2 }}\label{eqn:random walk gaussian distribution}
\end{align}
where $\sigma$ is given by,
\begin{align}
	\sigma &= \sqrt{\frac{1}{2}\sum_{n=1}^{\infty} c_n^2} \label{eqn: position space sigma}
\end{align}

An estimate for $\mathcal{E}_{HO}(q)$ can now be given by,
\begin{align}\label{eqn: eho in terms of mu and sigma}
	\mathcal{E}_{HO}(q) = c_0 + \sigma
\end{align}

For our case, from \autoref{eqn:delta tau for important paths}, we have,
\begin{align}\label{eqn: c0 definition}
	c_0 &= \frac{1}{\beta}\frac{1}{\sqrt{\alpha_0 + \frac{1}{2 \sum_{n=1}^\infty  \frac{1}{\alpha_n}}}}\\
	c_{n\geq1} &= \frac{1}{\beta}\frac{2}{\sqrt{\alpha_n}}\label{eqn: cn definition}
\end{align}

Let us first simplify the following expression,
\begin{align}
	\frac{1}{\alpha_0 + \frac{1}{2 \sum_{n=1}^{\infty}  \frac{1}{\alpha_n}}} &= \frac{1}{\alpha_0 + \frac{1}{\beta^2\braket{\Hat{q}^2} - \frac{1}{\alpha_0}}} \qquad \text{from \autoref{eqn:sum_as_q2}}\\
	&= \frac{\beta^2\braket{\Hat{q}^2} - \frac{1}{\alpha_0}}{\alpha_0 \beta^2\braket{\Hat{q}^2} } \label{eqn:c_0_simplification}
\end{align}

Hence, from \autoref{eqn: position space sigma}, \autoref{eqn: eho in terms of mu and sigma}, \autoref{eqn: c0 definition} and \autoref{eqn: cn definition}, we have,
\begin{align}
	\mathcal{E}_{HO}(q) &= \frac{1}{\beta} \left( \sqrt{\frac{\beta^2\braket{\Hat{q}^2} - \frac{1}{\alpha_0}}{\alpha_0 \beta^2\braket{\Hat{q}^2} }} + \sqrt{\beta^2\braket{\Hat{q}^2} - \frac{1}{\alpha_0}} \right) \qquad \text{from \autoref{eqn:sum_as_q2}}\\
    &=  \sqrt{ \braket{\Hat{q}^2}  - \frac{1 }{\beta^2 \alpha_0}} \left( \sqrt{\frac{1 }{\alpha_0 \beta^2\braket{\Hat{q}^2} }} + 1 \right)\\
    &= \sqrt{ \braket{\Hat{q}^2}  - \frac{1 }{m \omega^2 \beta}} \left( \sqrt{\frac{1 }{m \omega^2 \beta\braket{\Hat{q}^2} }} + 1 \right) \label{eqn:quantum_deviation_harmonic_oscillator}
\end{align}
where we have used $\alpha_0 = \frac{m \omega^2}{\beta}$ from \autoref{eqn:expression_for_alpha_n}. This recovers the expression for $\mathcal{E}(q)$ [\autoref{eqn:quantum_deviation_arbitrary_potential}], the typical quantum deviation of all the paths about $q_{cl}(\tau)$ that have a significant contribution to the action.

\section{Special cases} \label{Appendix:special cases}

\subsection{Ultra-strong coupling limit}\label{Appendix:Ultra-strong coupling limit}

In the ultra-strong coupling limit, i.e., $\lim_{J(\omega) \to \infty}$, first note that, from the definition of $\zeta_n$ [\autoref{eqn:HO_zeta}], we have,
\begin{align}\label{eqn: divergent zeta}
	\lim_{J(\omega) \to \infty} \zeta_{n \neq 0} = \infty
\end{align}

From the definition of $\braket{\hat{X}^2(q)}$ [\autoref{eqn:q^2 of the fitted oscillator}], we have,
\begin{align}
	\lim_{J(\omega) \to \infty} \braket{\hat{X}^2(q)} &= \lim_{J(\omega) \to \infty} \frac{1}{m \beta} \sum_{n=-\infty}^{+\infty} \frac{1}{\frac{V_2(q)}{m}+\nu_n^2+\zeta_n}\\
	&= \frac{1}{\beta V_2(q)}
\end{align}

Similarly, from definition of $\braket{\hat{P}^2(q)}$ [\autoref{eqn:p^2 of the fitted oscillator}], we have,
\begin{align}
	\lim_{J(\omega) \to \infty} \braket{\hat{P}^2(q)}
	&=  \frac{m}{\beta} + 2  \sum_{n=1}^{+\infty} \lim_{\zeta_n \to \infty} \frac{\frac{V_2(q)}{m} + \zeta_n}{\frac{V_2(q)}{m}+\nu_n^2+\zeta_n}\\
	&=  \frac{m}{\beta} + 2  \sum_{n=1}^{+\infty} 1\\
	&=  \infty
\end{align}

Finally, from definition of $\mathcal{G}(q)$ [\autoref{eqn:deviation integral arbitrary potential}], we have,
\begin{align}
	\lim_{J(\omega) \to \infty} \mathcal{G}(q) &= \lim_{J(\omega) \to \infty} \frac{1}{\sqrt{V_2(q) \beta \braket{\hat{X}^2(q)}}}\prod_{n=1}^\infty \frac{\nu_n^2 + \zeta_n }{\mathcal{A}_n(q)}\\
	&= \lim_{J(\omega) \to \infty} \frac{1}{\sqrt{V_2(q) \beta \frac{1}{\beta V_2(q)}}}\prod_{n=1}^\infty 1\\
	&= 1
\end{align}

Hence expression for USC-LHA [\autoref{eqn:HO_approx}] becomes,
\begin{align}
	\lim_{J(\omega) \to \infty}\rho(q,\eta) &= \delta(\eta) Z^{-1} \exp \left(- \frac{(V_1 (q)/V_2(q))^2}{2 \braket{\hat{X}^2(q)}} -\beta \left (V(q)-\frac{V_1(q)^2}{2 V_2(q)} \right) \right)\\
	\implies \lim_{J(\omega) \to \infty}\rho(q)&= Z^{-1}\exp \left (-\beta V(q)\right) \exp \left( -\frac{\beta V_1(q)^2}{2 V_2(q)} + \frac{\beta V_1(q)^2}{2 V_2(q)}\right ) \\
	&= Z^{-1}\exp \left (-\beta V(q)\right) \label{eqn:ultra_strong_MFGS}
\end{align}

which is the CV version of the USC result \cite{cresser2021weak}. To see this equivalence, note that the USC result is usually stated as,
\begin{equation}
	\rho = Z_{USC}^{-1} \exp{\left(-\beta \sum_n P_n H_s P_n\right)}
\end{equation}
where $P_n = \ket{x_n} \bra{x_n}$ are projection operators  on the non-degenerate eigenstates $\ket{x_n}$ of the system coupling operators $X$ and $ Z_{USC}$ is a normalization factor.

We note that in the CV limit,
\begin{align}
	\sum_n P_n H_s P_n &= \int dq \left ( \frac{\bra{q} \hat{P}^2\ket{q}}{2 m}  + V(q) \right) \ket{q}\bra{q}
\end{align}
Now, $\sum_q \bra{q} \hat{P}^2 \ket{q} \ket{q}\bra{q}$ is proportional to the identity operator, and hence will commute with the potential operator. Therefore, in the expression for the USC-MFGS, the contribution of the momentum operator can be absorbed in the trace, and we recover \autoref{eqn:ultra_strong_MFGS} as the USC result.

Now, let us calculate $\Delta(q,\eta)$ in this limit. Let us first evaluate $W(q)$ [\autoref{eqn:fitted_oscillating_sum}] that comes up in the definition of $\Delta(q,\eta)$ as,
\begin{align}
	\lim_{J(\omega) \to \infty} W(q) &= \frac{1}{\beta V_2(q)} +  \frac{1}{m \beta} \sum_{n=1}^{+\infty} \lim_{\zeta_n \to \infty} \frac{(-1)^n}{\frac{V_2(q)}{m}+\nu_n^2+\zeta_n}\\
	&= \frac{1}{\beta V_2(q)}
\end{align}

Hence, from the definition of $\Delta(q,\eta)$ [\autoref{eqn:classical_deviation_arbitrary_potential}], we have,
\begin{align}
	\lim_{J(\omega) \to \infty} \Delta(q,\eta) &=  \max \left\{ \left | \lim_{J(\omega) \to \infty}
	\frac{V_1(q_0)}{V_2(q_0)} \left ( 1 - \frac{W(q_0)}{\braket{\hat{X}^2(q_0)}} \right) \right |,
	\left | \eta \right |
	\right\}\\
	&=  \max \left\{ \left |
	\frac{V_1(q_0)}{V_2(q_0)} \left ( 1 - \frac{\frac{1}{\beta V_2(q)}}{\frac{1}{\beta V_2(q)}} \right) \right |,
	\left | \eta \right |
	\right\}\\
	&=  \max \left\{ 0 ,
	\left | \eta \right |
	\right\}\\
\end{align}

Similarly, let us calculate $\mathcal{E}(q)$ in this limit. From the definition of $\mathcal{E}(q)$ [\autoref{eqn:quantum_deviation_arbitrary_potential}], we have,
\begin{align}
	\lim_{J(\omega) \to \infty} \mathcal{E}(q) &= \lim_{J(\omega) \to \infty} \sqrt{ \braket{\hat{X}^2(q)}  - \frac{1 }{V_2(q) \beta}} \left( \sqrt{\frac{1 }{V_2(q) \beta \braket{\hat{X}^2(q)} }} + 1 \right)\\
	&=  \sqrt{ \frac{1 }{V_2(q) \beta}  - \frac{1 }{V_2(q) \beta}} \left( \sqrt{\frac{1 }{V_2(q) \frac{1 }{V_2(q) \beta} }} + 1 \right)\\
	&=  0
\end{align}

Hence, we note that both $\Delta(q,\eta)$ and $\mathcal{E}(q)$ have expected behavior in the USC limit.

\subsection{High temperature limit}\label{Appendix: High temperature limit}

To understand why LHA becomes accurate in the high temperature limit, consider an arbitrary path $q(\tau)$ starting and ending at the points $q+\eta$ and $q-\eta$, respectively. Now, for $a>1$, let us rescale $\beta \to \beta/a$ such that we have $q(\tau) \to q_a(\tau) \equiv q(a \tau)$, where we have identified the paths $q(\tau)$ and $q_a(\tau)$ with each other through a one-to-one and onto map. Note that we have $q(0) = q + \eta = q_a(0)$ and $q(\beta) = q - \eta = q_a(\beta/a)$. Now, define the new action, $S_a[q_a(\tau)]$ as,
\begin{align}\label{eqn:scaled action}
	S_a[q_a(\tau)] &= \int_0^{\beta/a} d\tau \left( \frac{1}{2} m \dot{q}_a(\tau)^2 + V\left(q_a(\tau)\right) \right) + \frac{1}{4}\int_0^{\beta/a} d\tau \int_0^{\beta/a} d\tau' K_a\left( \tau - \tau' \right) \left(q_a(\tau) - q_a(\tau')\right)^2
\end{align}
where,
\begin{equation}\label{eqn:scaled correlation function}
	K_a(\tau) = \int_0^\infty d\omega J(\omega) \frac{\cosh{[\omega(\beta/2a - \tau)]}}{\sinh{[\omega \beta/2a]}}
\end{equation}
Then, we have,
\begin{align}
    \int_0^{\beta/a} \frac{d^2}{d\tau^2}q_a(\tau) d\tau &= a^2 \int_0^{\beta/a} \frac{d^2}{d(a\tau)^2}q(a\tau) d\tau \\
    &= a \int_0^{\beta} \frac{d^2}{d\tau^2}q(\tau) d\tau
\end{align}
Similarly, we have,
\begin{align}
    \int_0^{\beta/a} V(q_a(\tau)) d\tau = \frac{1}{a} \int_0^{\beta} V(q(\tau)) d\tau
\end{align}

Before moving to the dissipative term, we write the hyperbolic factor coming in the correlation function [\autoref{eqn:scaled correlation function}] in the large $a$ limit for $0 \leq \tau \leq \beta/a$ as,
\begin{align}
    \lim_{a>>1} \frac{\cosh{[\omega(\beta/2a - \tau)]}}{\sinh{[\omega \beta/2a]}} = \frac{2a}{\omega \beta}
\end{align}
For the dissipative term, $T_D$, in \autoref{eqn:scaled action}, we then have,
\begin{align}
    T_D &= \lim_{a>>1} \int_0^{\beta/a}d\tau' \int_0^{\beta/a}d\tau  K_{a}(\tau - \tau') (q_a(\tau) - q_a(\tau'))^2\\
    &= \lim_{a>>1} \int_0^\infty d\omega J(\omega) \int_0^{\beta/a}d\tau' \int_0^{\beta/a}d\tau  \frac{\cosh{[\omega(\beta/2a - (\tau - \tau'))]}}{\sinh{[\omega \beta/2a]}} (q(a\tau) - q(a\tau'))^2\\
    &\approx \lim_{a>>1} \int_0^\infty d\omega J(\omega) \int_0^{\beta/a}d\tau' \int_0^{\beta/a}d\tau  \frac{2a}{\omega \beta} (q(a\tau) - q(a\tau'))^2\\
    &\approx \lim_{a>>1} \frac{1}{a} \int_0^\infty d\omega J(\omega) \int_0^{\beta}d\tau' \int_0^{\beta}d\tau  \frac{2}{\omega \beta} (q(\tau) - q(\tau'))^2
\end{align}
Hence, for large $a$, while the potential and dissipative terms scale as $1/a$, the kinetic term scales as $a$, dominating in the high temperature limit and causing the paths with large variation to contribute less.

Analogous to the USC limit [Appendix.~\ref{Appendix:Ultra-strong coupling limit}], the high temperature limit gives us the following values for the relevant quantities,
\begin{align}
    \lim_{\beta \to 0} \braket{\hat{X}^2(q)} &= \frac{1}{\beta V_2(q)}\\
    \lim_{\beta \to 0} \braket{\hat{P}^2(q)} &= \frac{m}{\beta}\\
    \lim_{\beta \to 0} \mathcal{G}(q) &= 1
\end{align}

Hence, the high temperature limit for the expression of LHA [\autoref{eqn:HO_approx}] is given as,
\begin{align}
    \lim_{\beta \to 0} \rho(q,\eta) &= \lim_{\beta \to 0} Z^{-1} \exp \left( - \frac{\beta V_2(q)}{2}\left( \frac{V_1 (q)}{V_2(q) } \right)^2  - \frac{m}{2\beta} \eta^2 -\beta \left( V(q)-\frac{V_1 (q)^2}{2 V_2(q)} \right)  \right)\\
    \implies \lim_{\beta \to 0} \rho(q) &= Z^{-1} \exp \left( - \beta V(q) \right) \label{eqn: high temperature mfgs appendix}
\end{align}

We will now show that this high-temperature MFGS is equivalent to the ones recently derived in literature \cite{cresser2021weak, gelzinis2020analytical, timofeev2022hamiltonian}.
Timofeev et al.\cite{timofeev2022hamiltonian} have shown that in the high temperature limit, at arbitrary coupling, the general Hamiltonian of mean force, $H_{MF}$, is given as,
\begin{align}\label{eqn: Timofeev high temperature hamiltonian of mean force}
	H_{MF} = \sum_n \langle n|H_S|n\rangle + \sum_{n\neq m} \bra{n}H_S\ket{m} \exp\{-\beta \Lambda (q_n - q_m)^2/6\} \ket{n} \bra{m}
\end{align}
where $\Lambda \equiv \int_0^\infty d\omega J(\omega)/\omega$ and $q_n$ are the eigenvalues of the system coupling operator $\Hat{q}$. Note that \autoref{eqn: Timofeev high temperature hamiltonian of mean force} differs from the one obtained by Timofeev et al. by a term of $-\Lambda \Hat{q}^2$ because their definition of the free Hamiltonian includes the counterterm. For a CV system, \autoref{eqn: Timofeev high temperature hamiltonian of mean force} translates into,
\begin{align}
	H_{MF} = \left( \int dx  V(x) \ket{x}\bra{x} \right) + \left( \int \int dx dy  \frac{\bra{x}\Hat{P}^2\ket{y}}{2m} \exp\left\{-\beta \Lambda \frac{(x - y)^2}{6}\right\} \ket{y}\bra{x} \right)\label{eqn:high T CV hamiltonian of mean force}
\end{align}

Let us first set $\Lambda = 0$ for simplicity. Then, we have,
\begin{align}
	\implies \exp\{-\beta \Hat{H}_{MF}\} &\approx \exp\{-\beta V(x)\} \times  \exp\left\{ -\beta \frac{\Hat{P}^2}{2 m} \right\} + O(\beta^2)
\end{align}
Let us focus on the second term,
\begin{align}
	\bra{x}\exp(-\beta \Hat{P}^2/2m) \ket{y} &= \sqrt{\frac{m}{2\pi \beta}} \exp\left\{- \frac{m}{2\beta} (x-y)^2\right\}\\
	&\approx \delta(x-y) \qquad \text{in high temperature limit}
\end{align}
We hence recover the high temperature MFGS result [\autoref{eqn: high temperature mfgs appendix}]. Note that when $\Lambda \neq 0$, then the off-diagonal elements of $H_{MF}$ [\autoref{eqn:high T CV hamiltonian of mean force}] get further suppressed and it can be formally shown that the exponential of the second term in \autoref{eqn:high T CV hamiltonian of mean force} gets closer to $\delta(x-y)$ as compared with the $\Lambda=0$ case, and hence \autoref{eqn: high temperature mfgs appendix} remains the high temperature MFGS.

We will now calculate $\Delta(q,\eta)$ [\autoref{eqn:classical_deviation_arbitrary_potential}] and $\mathcal{E}(q)$ [\autoref{eqn:quantum_deviation_arbitrary_potential}] in the high temperature limit. For $\Delta(q,\eta)$, we first evaluate,
\begin{align}
	\lim_{\beta \to 0} \frac{V_1(q)}{V_2(q)} \left ( 1 - \frac{W(q)}{\braket{\hat{X}^2(q)}} \right) &= \lim_{\beta \to 0} \frac{V_1(q)}{V_2(q)} \left ( 1 - \frac{\sum_{n=-\infty}^{+\infty} \frac{(-1)^n}{\frac{V_2(q)}{M}+\nu_n^2+\zeta_n}}{\sum_{n=-\infty}^{+\infty} \frac{1}{\frac{V_2(q)}{M}+\nu_n^2+\zeta_n}} \right)\\
    &= \lim_{\beta \to 0} \frac{V_1(q)}{V_2(q)} \left ( 1 - \frac{\frac{M}{V_2(q)}}{\frac{M}{V_2(q)}} \right)\\
    &= 0
\end{align}
We then have,
\begin{align}
    \lim_{\beta \to 0} \Delta(q,\eta) = \left| \eta \right|
\end{align}

Similarly, for $\mathcal{E}(q)$, we have,
\begin{align}
    \lim_{\beta \to 0} \mathcal{E} &= \lim_{\beta \to 0} \sqrt{ \braket{\Hat{q}^2}  - \frac{1 }{m \omega^2 \beta}} \left( \sqrt{\frac{1 }{m \omega^2 \beta \braket{\Hat{q}^2} }} + 1 \right)\\
    &= 2 \lim_{\beta \to 0} \sqrt{ \frac{2}{m \beta} \sum_{n=1}^\infty \frac{1}{\omega^2 + \nu_n^2 + \zeta_n}}\\
    &= 0
\end{align}

Since these length scales become smaller as the temperature is increased, we infer that LHA becomes more accurate in the process.

\subsection{High bath cutoff limit}
Let $\Omega$ denote the bath cutoff frequency. Then in the limit $\Omega \to \infty$, we have,
\begin{align}
    \lim_{\Omega \to \infty} \int_0^\infty \frac{J(\omega)}{\omega} d\omega = \infty
\end{align}

From the definition of $\zeta_n$ [\autoref{eqn:HO_zeta}], we have,
\begin{align}
	\lim_{n\to \infty} \zeta_n &= \frac{1}{m} \int _{0}^{\infty} \frac{J(\omega')}{\omega'} d\omega'\\
	&= \infty\\
	\implies \lim_{n \to \infty} \frac{\omega^2 + \zeta_n}{\omega^2 + \nu_n ^2 +  \zeta_n} &= 1\\
	\implies \lim_{n \to \infty} \sum_{n=1}^\infty \frac{\omega^2 + \zeta_n}{\omega^2 + \nu_n ^2 +  \zeta_n} &= \infty\\
	\implies \braket{\hat{P}^2(q)} &= \infty
\end{align}
Here we have used the definition of $\braket{\hat{P}^2(q)}$ [\autoref{eqn:p^2 of the fitted oscillator}]. Since $\braket{\hat{P}^2(q)}$ diverges, in this limit, LHA [\autoref{eqn:HO_approx}] will not have coherence in the position basis, and therefore will be given by,
\begin{equation}
    \rho(q,\eta) = \delta(\eta) Z^{-1} \mathcal{G}(q) \exp \left(- \frac{1}{2 \braket{\hat{X}^2(q)}} \left( \frac{V_1 (q)}{V_2(q)}\right)^2 -\beta \left(V{(q)}-\frac{V_1 {(q)}^2}{2 V_2{(q)}} \right) \right)
\end{equation}

\end{appendices}
\end{widetext}
\end{document}